\documentclass[10pt,a4paper]{article}
\usepackage{epsfig}
\usepackage{amstex}
\usepackage{amssymb}
\usepackage[square]{natbib}
\newcommand{\rf}[1]{(\ref{#1})}
\newcommand{\scaption}[1] {\caption{\footnotesize{#1}}}

\newcommand{\smi}  {\mbox{\scriptsize{i}}}
\begin{document}
\title{Baroclinic flow and the Lorenz-84 model}
\author{Lennaert van Veen\footnote{Mathematical Institute, University of Utrecht, PO Box 80.010, 3508 TA Utrecht, the Netherlands}}
\maketitle

\begin{abstract}
The bifurcation diagram of a truncation to six degrees of freedom of the equations for quasi-geostrophic,
baroclinic flow is investigated. Period doubling cascades and Shil'nikov bifurcations lead to chaos in
this model. The low dimension of the chaotic attractor suggests the possibility to reduce the model
to three degrees of freedom. In a physically comprehensible limit of the parameters this reduction
is done explicitly. The bifurcation diagram of the reduced model in this limit
is compared to the diagram of the six degrees of freedom model and agrees well.
A numerical implementation of the graph transform is used
to approximate the three dimensional invariant manifold away from the limit case. 
If the six dimensional
model is reduced to a linearisation of the invariant manifold about the Hadley state, the Lorenz-84
model is found. Its parameters can then be calculated from the physical parameters of the quasi-geostrophic
model. Bifurcation diagrams at physical and traditional parameter values are compared
and routes to chaos in the Lorenz-84 model are described.
\end{abstract}

\section{Introduction}

The equations for atmospheric flow form one of the most intensely studied dynamical systems of the last century.
Both their practical importance and their mathematical richness have attracted much attention. The atmospheric 
equations are studied in various forms, depending on the physical domain, and the time and length scales of
interest. The starting point of the analysis in this paper is a model which describes midlatitude atmospheric flow 
on a synoptic scale, a few thousand kilometers in space and a week or so in time.
A phenomenological description of the typical flow patterns in this range can be found in \cite{peix}, chapter 7.
The model consists of the filtered equations, derived from the basic atmospheric equations under the 
assumption of quasi-geostrophic (QG) and hydrostatic balance (see \cite{holt}).

In the absence of dissipative processes and forcing through solar heating, 
the filtered equations are energy preserving. The dominant dissipative terms are friction at the earths surface,
internal friction and Newtonian cooling. The solar heating induces a strong temperature gradient in the meridional
direction. Additional temperature gradients in the zonal direction can be induced by, e.g. land-sea contrast.
The response to this forcing is a strong westerly circulation, called the jet stream. This circulation can
become dynamically unstable so that traveling waves develop.

The jet stream pattern is nearly equivalent barotropic, which means that its height dependence can be represented by 
multiplication by a scale function. In other words, at each surface of constant height,
the velocity field has the same shape, but may have a different amplitude. In contrast, the traveling waves
can be baroclinic, which means that their phase depends on the vertical coordinate.
Typically, they exhibit a westward tilt with height. Theoretical studies of the filtered equations indicate, that
the baroclinicity of these traveling waves changes in the course of their life cycle
\cite[]{fris}. In the growing, strongly baroclinic, phase, they extract energy from the jet stream. In the decaying
phase, they can become equivalent barotropic and transfer energy back into the jet stream.

The focus of this study is on the interaction of the jet stream and the baroclinic waves and its representation
in a low order model. With the aid of discretisation
in the vertical and Galerkin truncation in the horizontal coordinates, we approximate the filtered equations
by a finite number of ordinary differential equations (ODE's). The discretisation can be done without violating
the energy preserving nature of the filtered equations, as described in \cite{lor4}. The number of layers is
fixed to two, the minimal number necessary to describe baroclinic waves.

The two layer model is considered on an $f$-plane, a rectangular domain on which the Coriolis force is taken
to be constant. The partial differential equations of the two layer model
are then projected onto Fourier modes. In each layer, we use one zonally symmetric 
pattern, representing the jet stream, and two patterns which combine to represent a traveling wave.
Thus, the ODE model has six degrees of freedom. The solar forcing is represented by constant terms, which are used
as bifurcation parameters. 

The bifurcation diagram is organised by its codimension two points, namely
fold-Hopf, 2:1 resonance and a neutral saddle-focus on a homoclinic bifurcation line.
Two routes to chaos are readily identified: period doubling cascades and a Shil'nikov type bifurcation.
An inspection of the spectra of equilibria
and periodic orbits found, and the calculation of the Kaplan-Yorke dimension of the chaotic attractor, leads
to the conjecture, that there is a three dimensional, globally attracting, invariant manifold in the phase space 
of the six dimensional model.

In order to calculate a first approximation of this invariant manifold, a small parameter is introduced
into the equations. In the limit where this parameter tends to zero, an analytic expression is obtained.
This limit has a clear physical interpretation in terms of the energy transfer between the jet stream
pattern and the traveling waves. Numerical evidence for the persistence of the invariant manifold away
from this limit is obtained by the use of techniques introduced by \cite{bro1} and \cite{foia}.

Reducing the six dimensional model to the invariant manifold, in the limit where the small parameter
tends to zero, a three dimensional model is obtained. 
Its bifurcation diagram 
is compared to the corresponding diagram of the six dimensional model in order to see if the qualitative
dynamics is retained. This comparison is convincing. Particularly, the codimension two points are
still present.

If the six dimensional model is reduced to a linearisation of the invariant manifold about the
Hadley state, the
Lorenz-84 model emerges. This model was introduced by \cite{lor2}
as the simplest model capable of representing the basic features of midlatitude, synoptic flow.
To the author's knowledge, no 
derivation of the Lorenz-84 model from atmospheric flow equations has been presented before. 
A rather {\sl ad hoc} link was established by \cite{wiin}, but in his work the reduction to three
degrees of freedom is not based on physical or mathematical arguments.
The link established
here enables us to calculate the parameters in the Lorenz-84 model from the physical parameters in the 
filtered equations. As it turns out, one of the parameters comes out significantly different from its traditional, 
yet unmotivated, value. A continuation in this parameter relates the bifurcation
diagram found at the traditional parameter value, presented in \cite{shil}, to the one found at the physical value.
The latter still bears resemblance to the bifurcation diagram of the six dimensional model, but the 
neutral saddle-focus transition is no longer there. Hence the route to chaos through a Shil'nikov type
bifurcation is absent. It is shown, that chaos through period doubling cascades, the Ruelle-Takens scenario
and intermittency does occur in the Lorenz-84 model.

The derivation presented here is not unlike the derivation of the Lorenz-63 model from the fluid dynamical
equations governing Rayleigh-B\'enard convection \cite[]{sal1,lor6}. There too, the Galerkin truncation
calculated has six degrees of freedom and can be reduced to a three dimensional invariant manifold. 
In that case, however, the invariant manifold is linear.

Galerkin truncations of the filtered equations have been
studied on various domains and at various truncation numbers, see \cite{swa2} and references therein. One lesson
to be learned from these studies, is that severe truncations, such as the one studied here, can only be regarded 
as qualitative models. A quantitative comparison to solutions of the filtered equations may be sensible at a 
truncation number in the order of a hundred or higher, depending on the basis functions used (see, e.g., 
\cite{ach2} and \cite{itoh}). 
Low order models, however, allow us to isolate a physical process
, such as the interaction between the jet stream and baroclinic waves, and represent it in a simple way.
Because of this conceptual simplicity, and the fact that they are easy to integrate numerically, 
low dimensional truncations are widely used for testing and illustrating new ideas in dynamical systems
theory, meteorology and climatology. The Lorenz-84 model for instance, has been used to investigate low-frequency
atmospheric variability \cite[]{piel}, measures of predictability \cite[]{gonz} and time scale interaction in
the climate system \cite[]{roeb,vee2}. The link with the filtered equations, presented here, validates the 
use of the Lorenz-84 model in these contexts, though with different parameter values. 

\section{The QG two layer model}

As a starting point for the calculations we take the quasi-geostrophic two layer model, described by \cite{lor4}. 
Alternative derivations of this model can be found e.g. in \cite{holt}, chapter 8, or the review article by 
\cite{swa2}. In Lorenz' article the stress is on the energy conserving nature of the nonlinear interaction terms, 
in De Swarts review article strong scaling arguments are provided.

\subsection{Setup of the model}

In the quasi-geostrophic approximation, the dry atmosphere is described by the velocity field, $\mathbf{v}$, and the 
temperature field, $T$. It is convenient to use the streamfunction, $\Psi$, the velocity potential, $\chi$, and the 
potential temperature, $\Theta$, as variables. As a vertical coordinate we use pressure instead of height.
The velocity and the temperature can then be expressed as
\begin{alignat}{2}
\mathbf{v}_{r} &= \mathbf{k}\times \nabla \Psi & \qquad \mathbf{v}_{d} &= \nabla \chi \nonumber \\
\mathbf{v} &= \mathbf{v}_{r}+\mathbf{v}_{d}+\omega \mathbf{k} & \qquad T &=\Theta \left( \frac{p}{p_{s}} \right)^{\frac{c_{p}-c_{v}}{c_{p}}} 
\label{vTPsiTheta}
\end{alignat}
Here $\mathbf{v}_{r}$ and $\mathbf{v}_{d}$ are the divergence free and the irrotational part of the horizontal 
velocity, $p$ is pressure, $p_{s}$ is surface pressure, $\omega=\mbox{d}p/\mbox{d}t$ is the vertical velocity,
$\mathbf{k}$ is the 
vertical unit vector and $c_{v}$ and $c_{p}$ are the specific heat of dry air at constant volume and pressure, 
respectively.

At the earth's surface, the lower boundary, we impose 
that $p=p_{s}$ is constant and $\omega=0$. At the upper boundary we have $p=0$ and $\omega=0$. Discretisation of 
the vertical in layers means that we replace each function of three spatial variables by a number of functions of 
longitude and latitude only. In the simplest case we take two layers, the minimal number necessary to describe 
baroclinic waves.

\begin{figure}[t]
\begin{picture}(350,185)
\centerline{\epsfig{file=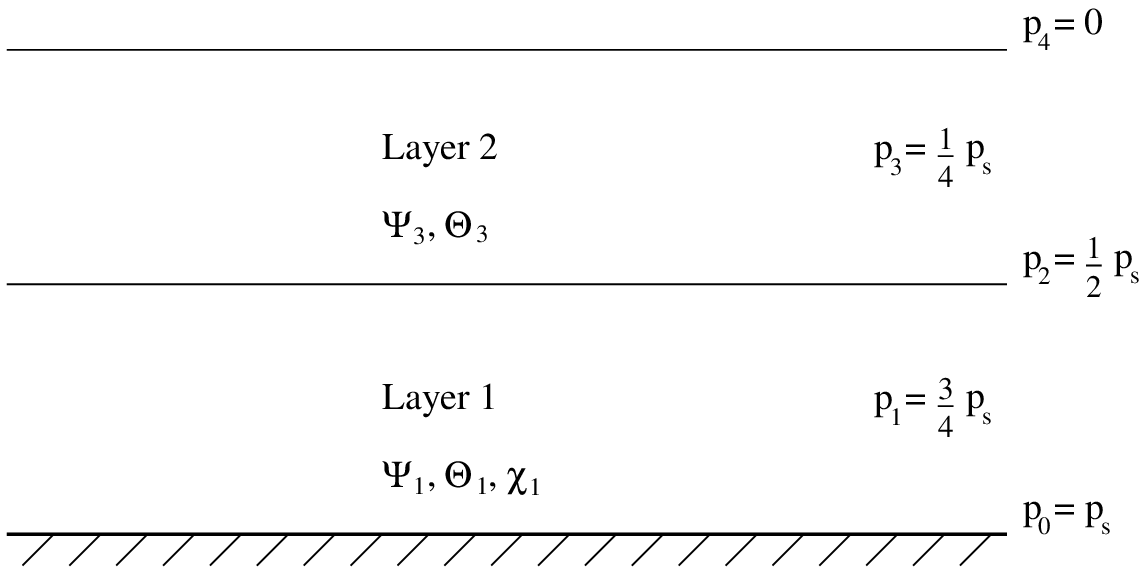,height=150pt}}
\end{picture}
\scaption{The vertical discretisation in pressure coordinates. If the static stability, $\sigma$,
is fixed, equations \rf{PDEsys1} and \rf{thermalwind} determine $\Psi$, $\tau$, $\Theta$ and $\chi_{1}$.}
\label{layers}
\end{figure}

The two layers are bounded by three isobaric surfaces at $p_{0}=p_{s}$, $p_{2}=p_{s}/2$ and $p_{4}=0$ (see figure~\ref{layers}). Vertical derivatives are replaced by linear interpolations, e.g. in the continuity equation:
\begin{equation}
\nabla^{2} \chi + \frac{\partial \omega}{\partial p} = 0 \: \rightarrow \: \left\{ \begin{array}{l}
\nabla^{2} \chi_{3} + \omega(p_{2}) =0 \\
\nabla^{2} \chi_{1} - \omega(p_{2}) =0 \end{array} \right.
\: \Rightarrow \: \nabla^{2}(\chi_{1}+\chi_{3}) =0
\end{equation}
The streamfunction and the potential temperature in the lower and the upper layer are denoted by $\Psi_{1,3}$ and 
$\Theta_{1,3}$, respectively. The pressure in the lower 
and the upper layer is set to $p_{1}=3p_{s}/4$ and $p_{3}=p_{s}/4$. The equations are written in terms 
of vertical means and differences, defined as
\begin{alignat}{2}
\Psi &= 1/2(\Psi_{3}+\Psi_{1}) & \qquad  & \text{the barotropic streamfuction,} \nonumber \\
\tau &= 1/2(\Psi_{3}-\Psi_{1}) & \qquad  & \text{the baroclinic streamfuction,} \nonumber \\
\Theta &= 1/2(\Theta_{3}+\Theta_{1}) & \qquad  & \text{the mean potential temperature,} \nonumber \\
\sigma &= 1/2(\Theta_{3}-\Theta_{1}) & \qquad  & \text{the static stability.} \label{def}
\end{alignat}
The static stability, $\sigma$, will be taken constant.

In addition to the conservative dynamics described in \cite{lor4}, we introduce linear damping through friction at 
the earth's surface, the terms proportional to $C$, friction at the boundary of the two layers, the term 
proportional to $C'$ and Newtonian cooling, the term proportional to $h_{N}$. 
The temperature forcing is given by $\Theta^{*}$. The resulting 
equations are
\begin{align}\label{PDEsys1}
\frac{\partial}{\partial t} \nabla^{2} \Psi & = -J(\Psi,\nabla^{2}\Psi +f) - J(\tau,\nabla^{2} \tau) -C \nabla^2 (\Psi - \tau) \tag{\ref{PDEsys1}.1}\\
\frac{\partial}{\partial t} \nabla^{2} \tau & = -J(\tau,\nabla^{2}\Psi +f) - J(\Psi,\nabla^{2} \tau) +\nabla\cdot (f\nabla\chi_{1}) + C \nabla^2 (\Psi - \tau) -2C' \nabla^2 \tau \tag{\ref{PDEsys1}.2}\\
\frac{\partial}{\partial t}\Theta & = -J(\Psi,\Theta) + \sigma\nabla^{2}\chi_{1} - h_{N}(\Theta-\Theta^{*})  \tag{\ref{PDEsys1}.3}
\end{align}
\addtocounter{equation}{1}
where the {\em Jacobian operator}, $J$, is defined as $J(A,B)=\nabla A \cdot \, \nabla B \times \mathbf{k}$ for any 
pair of functions $A,B$. The Coriolis parameter has been denoted by $f$. 

Furthermore we have the thermal wind equation, relating the shear streamfunction to the mean potential temperature
\begin{equation}
bc_{p}\nabla^{2}\Theta=\nabla\cdot(f\nabla\tau)
\label{thermalwind}
\end{equation}
Where $b=\frac{1}{2} [\left(\frac{p_{1}}{p_{s}}\right)^{\frac{c_{p}-c_{v}}{c_{p}}} - \left(\frac{p_{3}}{p_{s}}\right)^{\frac{c_{p}-c_{v}}{c_{p}}}] \approx 0.124$ comes out of the discretisation scheme in the vertical.

\subsection{Domain and boundary conditions}
\label{bouncon}

The set of equations \rf{PDEsys1} will be considered on an $f$-plane, a rectangular domain centered about a 
fixed latitude $\phi_{0}$ on which the Coriolis parameter is approximated by the constant value $f_{0}$. This 
domain has  length $L$ in the zonal direction and $D$ in the meridional direction. On this plane we will use 
Cartesian coordinates $x\in[0,L)$ and $y\in[0,D]$. In the following we set $\phi_{0}=45^{\circ}$.

In the zonal direction we take periodic boundary conditions. In the meridional direction we have
\begin{alignat}{2} \label{bc}
\frac{\partial \Psi}{\partial x} = \frac{\partial \tau}{\partial x} = \frac{\partial \chi_{1,3}}{\partial y} &=0 & \qquad \text{at } y=0,D
\tag{\ref{bc}.1}
\end{alignat}
This means that there is no mass flux through the boundaries. The second condition was put forward by \cite{phil}, and imposes that there is no net flow along the boundaries.
\begin{alignat}{2}
\int_{0}^{L}\frac{\partial \Psi}{\partial y}\mbox{d}x = \int_{0}^{L}\frac{\partial \tau}{\partial y}\mbox{d}x &=0 & \qquad \text{at } y=0,D
\tag{\ref{bc}.2}
\end{alignat}
It follows from the thermal wind equation \rf{thermalwind}, and the restriction that there be no net heat flux through the boundaries, that $\Theta$ satisfies
\begin{alignat}{3}
\frac{\partial \Theta}{\partial x} &= 0 &\qquad \int_{0}^{L}\frac{\partial \Theta}{\partial y}\mbox{d}x &= 0 & \qquad \text{at } y=0,D
\tag{\ref{bc}.3}
\end{alignat}
\addtocounter{equation}{1}
With these boundary conditions we can consider $\Theta$ to describe deviations from the spatially averaged potential temperature.

\subsection{Scaling}
\label{scaling}

In table \rf{scales} the scales, suitable for the synoptic physics are listed. In the right column the numerical 
values, used below, are listed. In the following all quantities are dimensionless, unless otherwise indicated. The 
dimensionless length of the domain will be denoted by $s=L/D$.
\begin{table}[h]
\begin{center}
\begin{tabular}{|l|l|l|}
\hline
Length & $D$ & $5\cdot 10^{3}$km \\ \hline
Time & $\Sigma^{-1}$ & $7$ days \\ \hline
Temperature & $R=\frac{f_{0}\Sigma D^{2}}{b c_{p}}$ & $ 34.3 $K \\ \hline
Mass & $M=\frac{p_{0}D^{2}}{2g} \frac{f_{0}}{\Sigma}\frac{R}{\sigma}$ & $ 7.7\cdot 10^{18}$ kg \\
\hline
\end{tabular}
\end{center}
\scaption{Scaling for the synoptic physics described by \rf{PDEsys1}. In the right column the 
numerical values used.}
\label{scales}
\end{table}

\subsection{Equations for $\Psi$ and $\tau$}

Under the boundary conditions \rf{bc}, the thermal wind relation \rf{thermalwind} takes the simple form
$\Theta=\tau$. This identity can be used to eliminate the velocity potential $\chi_{1}$ from equations 
(\ref{PDEsys1}.2) and (\ref{PDEsys1}.3). This results in a closed set of prognostic equations for $\Psi$ and $\tau$
\begin{align}\label{PDEsys2}
\frac{\partial}{\partial t} \nabla^{2} \Psi &= -J(\Psi,\nabla^{2}\Psi) - J(\tau,\nabla^{2} \tau) -C\nabla^2 (\Psi - \tau) \tag{\ref{PDEsys2}.1} \\
\frac{\partial}{\partial t}(1-\alpha\nabla^{2})\tau &= -J(\Psi,\tau) + \alpha J(\tau,\nabla^{2}\Psi)+ \alpha J(\Psi,\nabla^{2}\tau) - h_{N}(\tau-\tau^{*}) \nonumber \\
 &  \mbox{\hspace{135pt}} -\alpha C\nabla^2 (\Psi - \tau) +2\alpha C' \nabla^2 \tau \tag{\ref{PDEsys2}.2}
\end{align}
\addtocounter{equation}{1}
where $\alpha=\sigma/f_{0}$ and $\tau^*=\Theta^*$. In this scaling $f_{0}^{-1}\approx 1.6 \cdot 10^{-2}$ is the 
Rossby number. We will study the lowest dimensional nontrivial spectral truncation of these equations.

\subsection{Energy}

In the absence of friction and forcing, the prognostic equations \rf{PDEsys2} conserve the 
sum of kinetic and available potential energy, defined respectively as
\begin{alignat}{3}
K &=\alpha\int_{0}^{1}\int_{0}^{s}(\nabla \Psi \cdot \nabla \Psi + \nabla \tau \cdot \nabla \tau)\mbox{d}x\mbox{d}y & \qquad A &= \int_{0}^{1}\int_{0}^{s} \Theta^{2}\mbox{d}x\mbox{d}y
\label{conserved1} \end{alignat}
in units $MD^2 \Sigma^2$. The simplified models will be shown to have a corresponding conserved quantity.

\section{The Galerkin approximation}
\label{galer}

In order to approximate equations \rf{PDEsys2} by a finite number of ODE's we do a 
Galerkin projection onto Fourier modes. 
On this basis the variables are given by
\begin{align}
\Psi(x,y,t) &= \sum_{n,m} \psi (m,n,t) \exp[\smi(mkx+nly)] \nonumber \\
\tau(x,y,t) &= \sum_{n,m} \theta (m,n,t) \exp[\smi(mkx+nly)]
\end{align}
where $k=2\pi/s$, $l=\pi$. The boundary conditions, and the restriction that $\Psi$ and $\tau$ are real variables 
impose that 
\begin{alignat}{3}
\psi(m,-n) &=-\psi(m,n) & \qquad  \theta(m,-n)&=-\theta(m,n) & \qquad \text{if } m & \neq 0 \nonumber \\
\psi(0,n) &= \psi(0,-n) & \qquad  \theta(0,n) &=\theta(0,-n) & & \nonumber \\
\psi(m,n) &=\psi^{*}(-m,-n) & \qquad \theta(m,n) &=\theta^{*}(-m,-n) & &
\end{alignat}
This Fourier decomposition is equivalent to the introduction of a basis of eigenfunctions of the Laplacian operator 
on our domain, with the specified boundary conditions. The eigenfunctions are $\phi_{0n}=\mbox{cos}nly$ for zonal 
wavenumber zero, and $\phi_{mn}=\mbox{e}^{\smi mkx}\mbox{sin}nly$ otherwise. 

If we apply the zonally symmetric forcing $\theta^*=\frac{1}{2}\Delta T \phi_{01}$, with temperature contrast 
$\Delta T$ between the boundaries, there is an exact solution to equations \rf{PDEsys2}. It is 
given by
\begin{equation}
\Psi=\tau=\frac{1}{2}\frac{h_{N}\Delta T}{h_{N}+2\alpha l^2 C'} \phi_{01}
\label{hadley}
\end{equation}
This solution is called the Hadley state and describes a strong jet in the upper layer, rising air at the south 
boundary and sinking air at the north boundary. The eigenfunctions with nonzero zonal wavenumber describe traveling 
waves which can grow if the Hadley circulation becomes dynamically unstable.

The projection of equations \rf{PDEsys2} is given by
\begin{multline} \label{trunc}\addtocounter{equation}{1}
\lambda_{cd}\dot{\psi}(c,d) = -\lambda_{cd} C (\psi(c,d) - \theta(c,d)) \nonumber \\
+kl\sum_{pqrs}\lambda_{rs}(ps-qr)\delta(p+r-c)\mu(q+s-d)\{\psi(r,s)\psi(p,q)+ \theta(r,s)\theta(p,q)\} 
\tag{\ref{trunc}.1} \end{multline} \vspace{-20pt}
\begin{multline} 
\bar{\lambda}_{cd}\dot{\theta}(c,d) = - h_{N}(\theta(c,d)-\theta^{*}(c,d))-\alpha \lambda_{cd} C (\psi(c,d) - \theta(c,d)) +2\alpha C' \lambda_{cd} \theta(c,d) \nonumber \\
+ kl\sum_{pqrs}(ps-qr)\delta(p+r-c)\mu(q+s-d) \{1-\alpha([p^2-r^2]k^2+[q^2-s^2]l^2)\}\psi(p,q)\theta(r,s) 
\tag{\ref{trunc}.2} \end{multline}
\addtocounter{equation}{-1}
where $\delta$ is the Kronecker delta, $\theta^*$ is the Fourier transform of $\tau^*$ and $\mu$ is defined as
\begin{equation}
\mu(a)=\int_{0}^{1} \mbox{e}^{\smi a l y}\mbox{d}y = 
\begin{cases} 
1 & \text{if $a=0$} \\
0 & \text{if $a$ is even} \\
\frac{-2}{\pi\smi a} & \text{if $a$ is odd}
\end{cases}
\end{equation}
The eigenvalues of the operators on the left hand side of equations \rf{PDEsys2} have been denoted 
by \mbox{$\lambda_{ab}=-(a^2 k^2+b^2 l^2)$} and \mbox{$\bar{\lambda}_{ab}=1-\alpha\lambda_{ab}$}.

\section{The six dimensional truncation}
\label{sixD}

The smallest nontrivial truncation of equations \rf{trunc} has six degrees of freedom. We set $s=2$ and define
\begin{alignat}{3}
x_{1} &= 2\psi(0,1) & \qquad y_{1} &= 2\theta(0,1) & \qquad  T_{1} &= 2\theta^{*}(0,1) \nonumber \\
x_{2} &= 2\sqrt{2}\mbox{Re}\psi(1,1) & \qquad y_{2} &= 2\sqrt{2}\mbox{Re}\theta(1,1) & \qquad T_{2} &= 2\sqrt{2}\mbox{Re}\theta^{*}(1,1) \nonumber \\
x_{3} &= 2\sqrt{2}\mbox{Im}\psi(1,1) & \qquad y_{3} &= 2\sqrt{2}\mbox{Im}\theta(1,1) & \qquad T_{3} &= 2\sqrt{2}\mbox{Im}\theta^{*}(1,1) \label{defs}
\end{alignat}
These variables satisfy the following equations
\begin{align} \label{odesys1}\dot{x}_{1} & = -C (x_{1} - y_{1})  \tag{\ref{odesys1}.1} \\
\lambda_{11}\dot{x}_{2} & = -\lambda_{11}C (x_{2} - y_{2}) + \lambda_{10}\delta (x_{1}x_{3}+ y_{1}y_{3}) \tag{\ref{odesys1}.2} \\
\lambda_{11}\dot{x}_{3} & = -\lambda_{11}C (x_{3} - y_{3}) - \lambda_{10}\delta (x_{1}x_{2}+ y_{1}y_{2}) \tag{\ref{odesys1}.3} \\
\bar{\lambda}_{01}\dot{y}_{1} & = -\alpha \lambda_{01} (C x_{1}- [C+2C'] y_{1}) - h_{N}(y_{1}-T_{1}) +\delta (x_{3}y_{2} - x_{2}y_{3}) \tag{\ref{odesys1}.4} \\
\bar{\lambda}_{11}\dot{y}_{2} & = -\alpha\lambda_{11} (C x_{2}- [C+2C'] y_{2}) - h_{N}(y_{2}-T_{2}) + \delta (\bar{\lambda}_{10} x_{1}y_{3}-\nu_{10} x_{3}y_{1}) \tag{\ref{odesys1}.5} \\
\bar{\lambda}_{11}\dot{y}_{3} & = -\alpha\lambda_{11} (C x_{3}- [C+2C']y_{3}) - h_{N} (y_{3}-T_{3}) -\delta (\bar{\lambda}_{10}x_{1}y_{2}-\nu_{10} x_{2}y_{1}) \tag{\ref{odesys1}.6}
\end{align}
\addtocounter{equation}{1}
where $\delta=8kl/(3\pi)$ and $\nu_{ab}=1+\alpha\lambda_{ab}$.
When we put dissipation and forcing to zero, the ODE system \rf{odesys1} has a conserved quantity $L$, defined by
\begin{equation}
L=-\alpha\lambda_{01}x_{1}^2-\alpha\lambda_{11}(x_{2}^2+x_{3}^2) +\bar{\lambda}_{01} y_{1}^2+\bar{\lambda}_{11}(y_{2}^2+y_{3}^2) \label{conserved2}
\end{equation}
which corresponds to the projection of the sum of kinetic and available potential energy defined in 
equation \rf{conserved1}. 
Using $L$ as a Lyapunov function we can show, that a trapping region for equations \rf{odesys1} is defined by
\begin{equation}
L \leq \frac{3 \| \mathbf{T}\|^{2}}{ (h_{N}-2\alpha\lambda_{11}C')^{2} } \label{trap}
\end{equation}
where $\| . \|$ denotes the $L_{2}$ norm, and we have assumed that $0< h_{N} \leq C$, $0< 2C'\leq C$ and 
$\sigma\leq1$, a realistic range for the parameters. Note, that the divergence of the vector field, defined
by equation \rf{odesys1}, is constant and negative. Therefore, volume elements always shrink.

\subsection{Bifurcation diagram}
\label{6dbif}

In figure \rf{diagram1} the partial bifurcation diagram of system \rf{odesys1} is shown. This diagram,
and all other diagrams below, have been calculated using the software package AUTO (\cite{doed}).
To obtain this picture, we varied 
$T_{1}$ and $T_{2}$, setting $T_{3}=0$. Due to a discrete symmetry in equations \rf{odesys1}, given by
\begin{alignat}{2}
\begin{array}{l}
\mathbf{x} \rightarrow \mathbf{x}'=\mathbf{R}\mathbf{x} \\
\mathbf{y} \rightarrow \mathbf{y}'=\mathbf{R}\mathbf{y} \\
\mathbf{T} \rightarrow \mathbf{T}'=\mathbf{R}\mathbf{T} \end{array} & \qquad \text{where }
\mathbf{R} &= \left( \begin{array}{lll}
1 & 0 & 0 \\ 
0 & 0 & -1 \\
0 & 1 & 0 \end{array} \right),
\end{alignat}
setting $T_{2}=0$ and varying $T_{3}$ yields the same diagram. This symmetry corresponds to a translation
$x\rightarrow x'=x+1/2$ in the physical domain. Furthermore, we have set $C=3.5$, $h_{N}=0.7$, $C'=0.5$ and $\sigma=0.9$.
This corresponds to a damping time scale of two days at the earth's surface and two weeks at the layer interface.
The thermal damping time scale is ten days and the temperature difference between the layers is about $34\mbox{K}$.

\begin{figure}[ht]
\begin{picture}(400,330)
\centerline{\epsfig{file=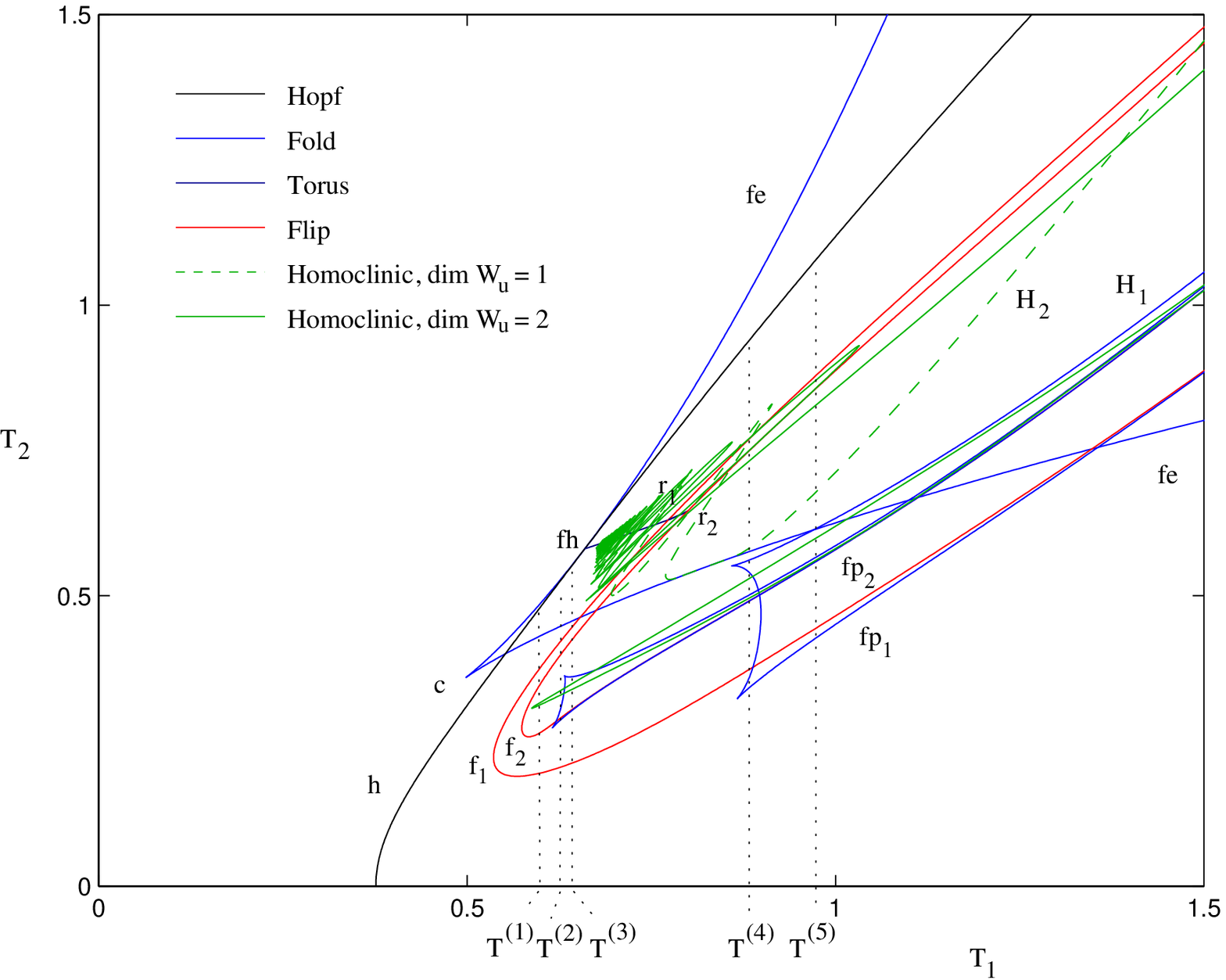,width=400pt}}
\end{picture}
\scaption{Bifurcation diagram of system \rf{odesys1}. The dotted lines at $T_{1}=T^{(1\ldots 5)}$ refer to figures 
\rf{seclim}-\rf{KYdim1}.}
\label{diagram1}
\end{figure}
If we put $T_{2}=0$ and increase $T_{1}$ from zero, at first a stable equilibrium is the unique limit set in the 
phase space of system \rf{odesys1}. This equilibrium corresponds to the Hadley state \rf{hadley}. At the Hopf 
bifurcation line, marked h, this equilibrium becomes unstable and a stable periodic orbit is created. This is our 
model's baroclinic instability. The periodic orbit corresponds to a traveling baroclinic wave.

The two line segments fe, joint at cusp point c, denote a fold bifurcation of the equilibrium. Within the \mbox{V-shaped} region, bounded 
by curve fe, there are three equilibria, one of which is stable. At the codimension two point, marked fh, the Hopf 
line and the fold line are tangent. At this point an equilibrium exists with one zero eigenvalue and a complex pair 
on the imaginary axis. The unfolding of this point can be found in \cite{kuz1}. In the following, we will adopt the notation of 
this book for normal form coefficients. An algorithm for computing the normal form coefficients of fold-Hopf
points is described in \cite{kuzn3} and implemented in the forthcoming release of CONTENT (\cite{kuzn4}).
In this case, we have normal form coefficients $s=1$ and $\theta<0$.

A torus bifurcation line emerges from point fh, and connects to the flip bifurcation line marked $\mbox{f}_{1}$. 
Above the torus bifurcation line, the periodic orbit created on Hopf curve h is unstable, below the torus bifurcation 
line it is stable and coexists with a saddle type, two dimensional torus. At the point where the torus and the flip 
bifurcation lines meet, the periodic orbit has two Floquet multipliers equal to minus unity. This point is marked 
$\mbox{r}_{1}$ for $1\!:\!2$ resonance. At this point, we have normal form coefficient $s=-1$. Connected to the 
resonance point $\mbox{r}_{1}$ there is torus bifurcation line of the period doubled orbit, which leads to another 
$1\!:\!2$ resonance point, $\mbox{r}_{2}$. In fact, flip bifurcation lines $\mbox{f}_{1}$ and $\mbox{f}_{2}$ are the 
first two of a period doubling cascade. There seems to be a accumulation of $1\!:\!2$ resonance points directly to 
the right of $\mbox{r}_{2}$. Such an accumulation has been described in the context of a biological model
in \cite{kuz1}, chapter 9.6. \cite{wiec} found it in a rate equation model for a
semiconductor laser. They also provide a partial unfolding of this codimension two phenomenon.

From codimension two point fh two homoclinic bifurcation lines emanate. Both homoclinic connections are
attached to a saddle focus. Along $\mbox{H}_{2}$ the unstable manifold of the saddle focus is one dimensional
and the saddle value, $\sigma$, is negative. On this line a stable periodic orbit is created.
Along $\mbox{H}_{1}$, the saddle focus has a complex pair of eigenvalues with positive real part.
The saddle value is positive along the larger part $\mbox{H}_{1}$ and another stable periodic
orbit is created. However, near the
leftmost turning point of this curve, the saddle value changes sign. On a small segment it is negative,
indicating that Shil'nikov type chaos can occur. In the literature, this type of Shil'nikov bifurcation,
with complex unstable leading eigenvalues, is uncommon. 

The points, where the saddle value is zero, are called
neutral saddle focus, or Belyakov, transitions. What is known about the unfolding of this transition
is summed up in \cite{cham}. In figure \rf{shil} the transition points
are shown in more detail, along with a phase portrait. An infinite number of cusps of fold lines of periodic
orbits are expected to accumulate here, corresponding to the creation of successive folds of the branch
of periodic solutions which becomes homoclinic on $\mbox{H}_{1}$. However, as described in \cite{glen},
the neutral saddle focus transition is continuous, and these cusp bifurcations correspond to orbits of
very high period which are hard to detect numerically. Looking at sections like in figure \rf{branchsw},
the case with negative saddle values cannot be discerned from the case with positive saddle value.

\begin{figure}
\begin{picture}(400,520)
\put(15,236){\epsfig{file=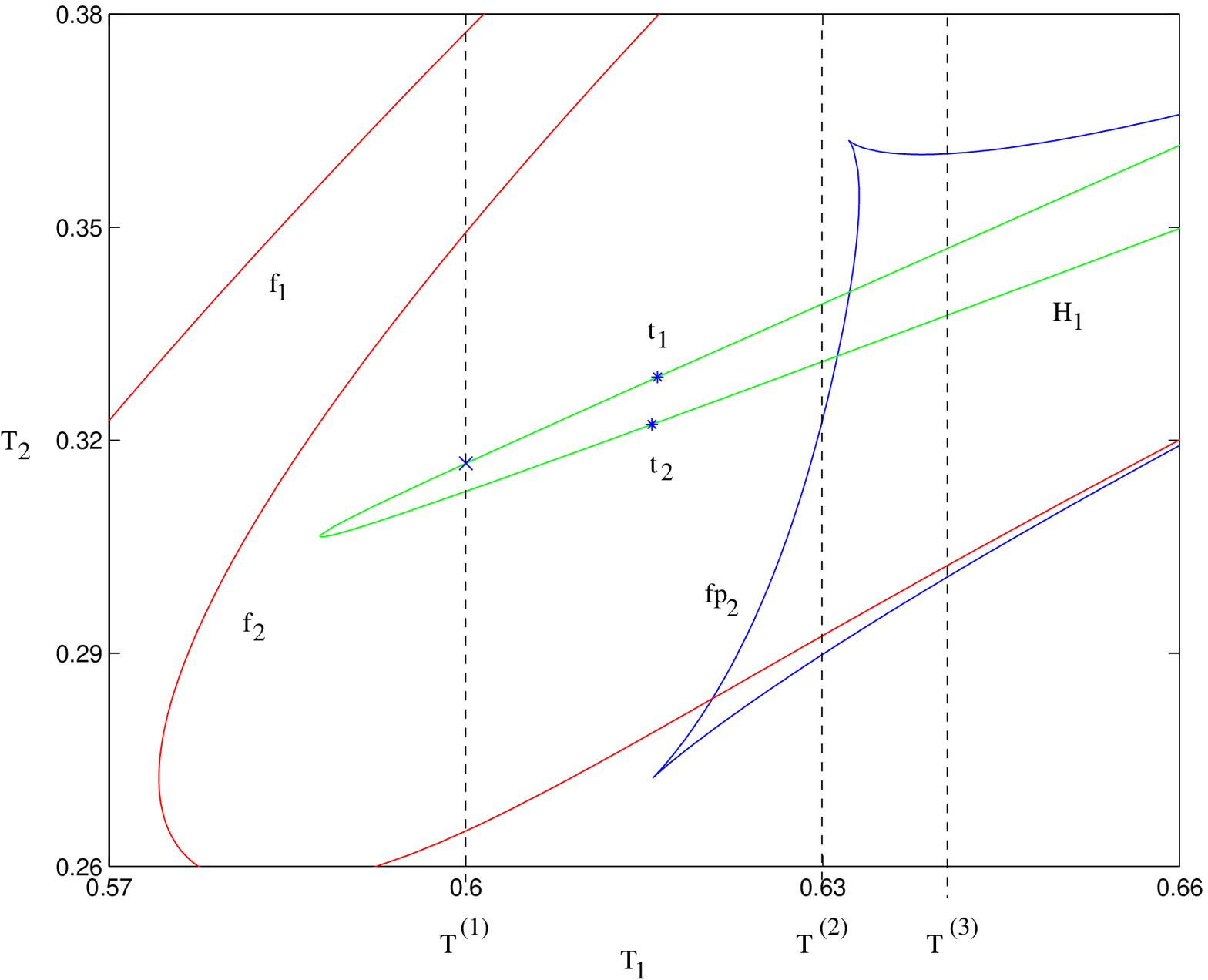,width=370pt}}
\put(15,-14){\epsfig{file=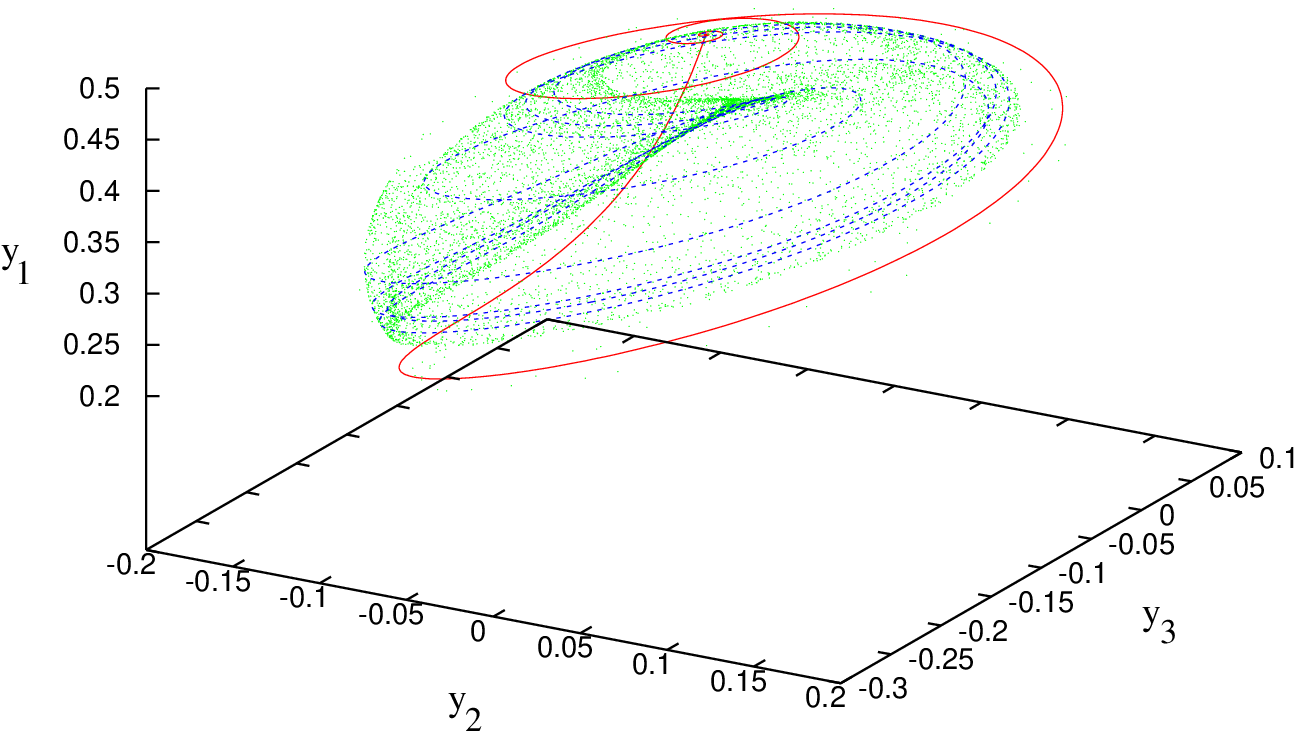,width=400pt}}
\end{picture}
\scaption{Top: detail of bifurcation diagram \rf{diagram1}. The neutral saddle focus transitions have been marked
$\mbox{t}_{1,2}$. To the left of these transitions the saddle value is negative and the Shil'nikov condition is
satisfied. Bottom: phase portrait at $(T_{1},T_{2})=(0.6 , 0.3168)$. These parameter values have been marked
with a cross in the top picture. In red: the homoclinic orbit. In blue, dashed: the periodic solution after
four period doublings, i.e. the fifth branch in the top picture of figure \rf{seclim}. In green: points on the chaotic attractor.}
\label{shil}
\end{figure}

In figure \rf{seclim} a cross section of diagram \rf{diagram1} is shown. We have fixed $T_{1}=T^{(1)}$ and let
$T_{2}$ vary as indicated in diagram \rf{diagram1}. The top picture shows a number of branches of the period
doubling cascade, as well as the primary homoclinic branch corresponding to curve $\mbox{H}_{1}$.
The bottom picture, on the same horizontal scale, is a limit point diagram. This picture was 
obtained by calculating the Poincar\'e map on the plane 
$\mathcal{S}=\{ (\mathbf{x},\mathbf{y})\in \mathbb{R}^{6} | y_{2}=0 \} $. After a sufficiently long integration, to 
filter out transients, the value of $y_{1}$ was plotted at a number of iterations of the Poincar\'e map. The 
behaviour is chaotic between the accumulation points of the period doubling cascade and changes qualitatively near 
the Shil'nikov bifurcation.

\begin{figure}
\begin{picture}(400,520)
\centerline{\epsfig{file=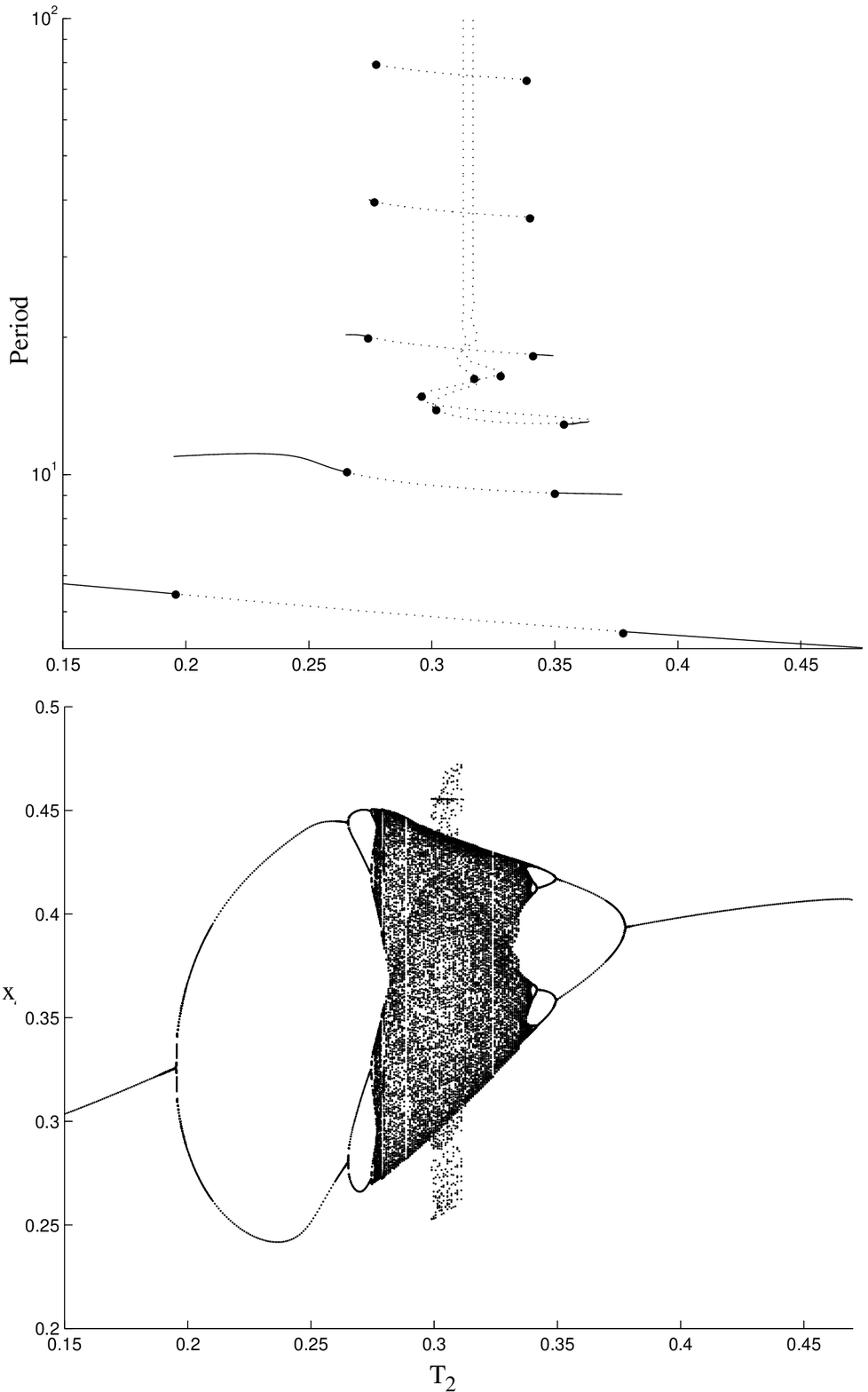,height=520pt}}
\end{picture}
\scaption{Top: continuation of the periodic orbit created on curve h, along the line $T_{1}=T^{(1)}$ 
in figure \rf{diagram1}. 
Solid lines denote stable branches, dotted lines denote unstable branches, Hopf bifurcation point are marked with 
dots. Also shown is the primary homoclinic branch corresponding to curve $\mbox{H}_{1}$. Bottom: limit point diagram 
of the Poincar\'e map on $\mathcal{S}$.}
\label{seclim}
\end{figure}

\begin{figure}[t]
\begin{picture}(415,358)
\put(0,0){\epsfig{file=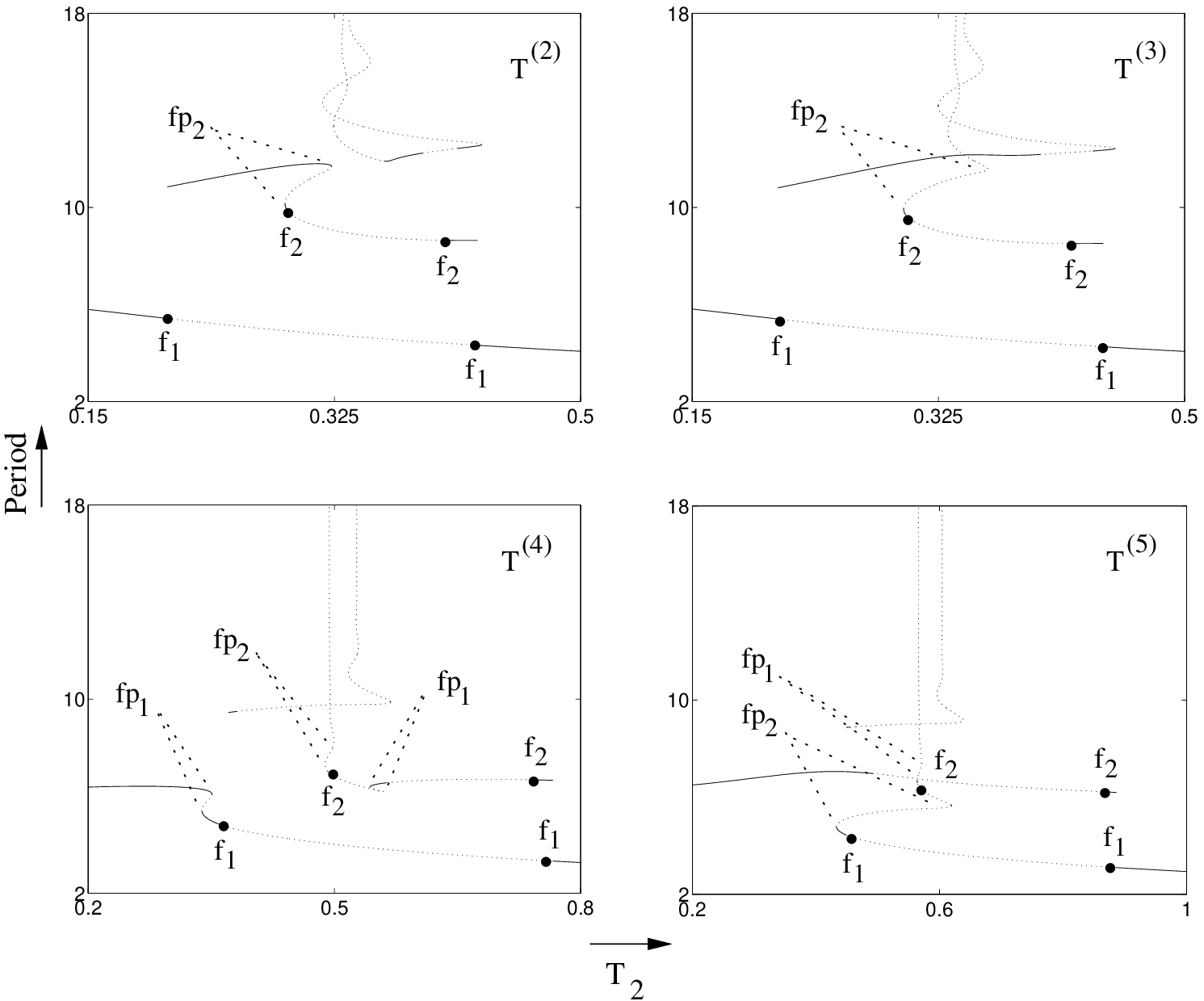,width=430pt}}
\end{picture}
\scaption{Cross sections of diagram \rf{diagram1}. Dots denote flip bifurcations, solid lines stable branches and
dotted lines unstable branches. The labels $\mbox{f}_{1,2}$
and $\mbox{fp}_{1,2}$ refer to diagram \rf{diagram1}. Top left: $T_{1}=T^{(2)}=0.63$. The basic cycle, born on
Hopf line h, and its period double version, are not connected to the homoclinic branch. Top right: $T_{1}=T^{(3)}=0.64$.
The period doubled cycle now becomes homoclinic on $\mbox{H}_{1}$. Bottom left: $T_{1}=T^{(4)}=0.88$. The branch 
of the basic cycle folds. Bottom right: $T_{1}=T^{(5)}=0.98$. From one end, $T_{2}$ increasing from zero, the
branch of the basic cycle ends in a period halving bifurcation. From the other end, $T_{2}$ decreasing, the
branch of the basic cycle becomes homoclinic on $\mbox{H}_{1}$.}
\label{branchsw}
\end{figure}
The lines marked $\mbox{fp}_{1}$ and $\mbox{fp}_{2}$ denote fold bifurcations of periodic orbits. Near the cusps of 
these fold lines, the periodic orbit, created at Hopf line h, and its period doubled versions, switch branches with 
periodic orbits that become homoclinic near $\mbox{H}_{1}$. This process is illustrated in figure \rf{branchsw}.
In the top left picture, where we have $T_{1}=T^{(2)}$, none of the branches of the period doubling cascade are
connected to the homoclinic branch, see also figure \rf{seclim}(top). For $T_{1}>T^{(2)}$, however, the first
period doubled branch becomes homoclinic on curve $\mbox{H}_{1}$. The period doubled branch and the homoclinic
branch collide in a transcritical bifurcation. This process is repeated for the branch of the basic cycle, continued
from Hopf line h. Therefore, the flip bifurcation lines $\mbox{f}_{1,2,\ldots}$ do not simply form a cascade and an
inverse cascade. At the fold lines $\mbox{fp}_{1,2}$ the simple structure of figure \rf{seclim}(top) at $T_{1}=T^{(1)}$
is rearranged.

Summarising, the qualitative dynamics of system \rf{odesys1} is as follows. Left of the Hopf line h, and within
the V-shaped region bounded by fold curve fe, there is a stable equilibrium. To the right of curve h, and below
curve fe, the behaviour is periodic before crossing flip bifurcation curve $\mbox{f}_{1}$. When crossing $\mbox{f}_{1}$
near the leftmost fold of the homoclinic curve $\mbox{H}_{1}$, where the saddle value is positive, a combination
of period doubling and Shil'nikov chaos is encountered, as demonstrated in figures \rf{shil} and \rf{seclim}.
To the right of the neutral saddle focus transition points, the behaviour is alternatingly periodic and chaotic.
Due to the branch switching, shown in figure \rf{branchsw}, the parameter space is divided into small chaotic and
periodic windows.

In figure \rf{KYdim1} the Kaplan-Yorke dimension of the attractor is shown for $T_{1}=T^{(1)}$ and values of $T_{2}$ at 
which complex dynamics arise. Most remarkably, the attractor dimension does not exceed three. Also, the equilibria 
and periodic orbits studied in figure \rf{diagram1} have a feature in common. They all have three strongly contracting 
directions. These observations suggest, that the dynamics of system \rf{odesys1} take place on a three dimensional 
invariant manifold. In the next section we will present an approximate, three dimensional invariant manifold of this 
system, which enables us to reduce the model to three degrees of freedom. 
\begin{figure}[h]
\begin{picture}(400,260)
\centerline{\epsfig{file=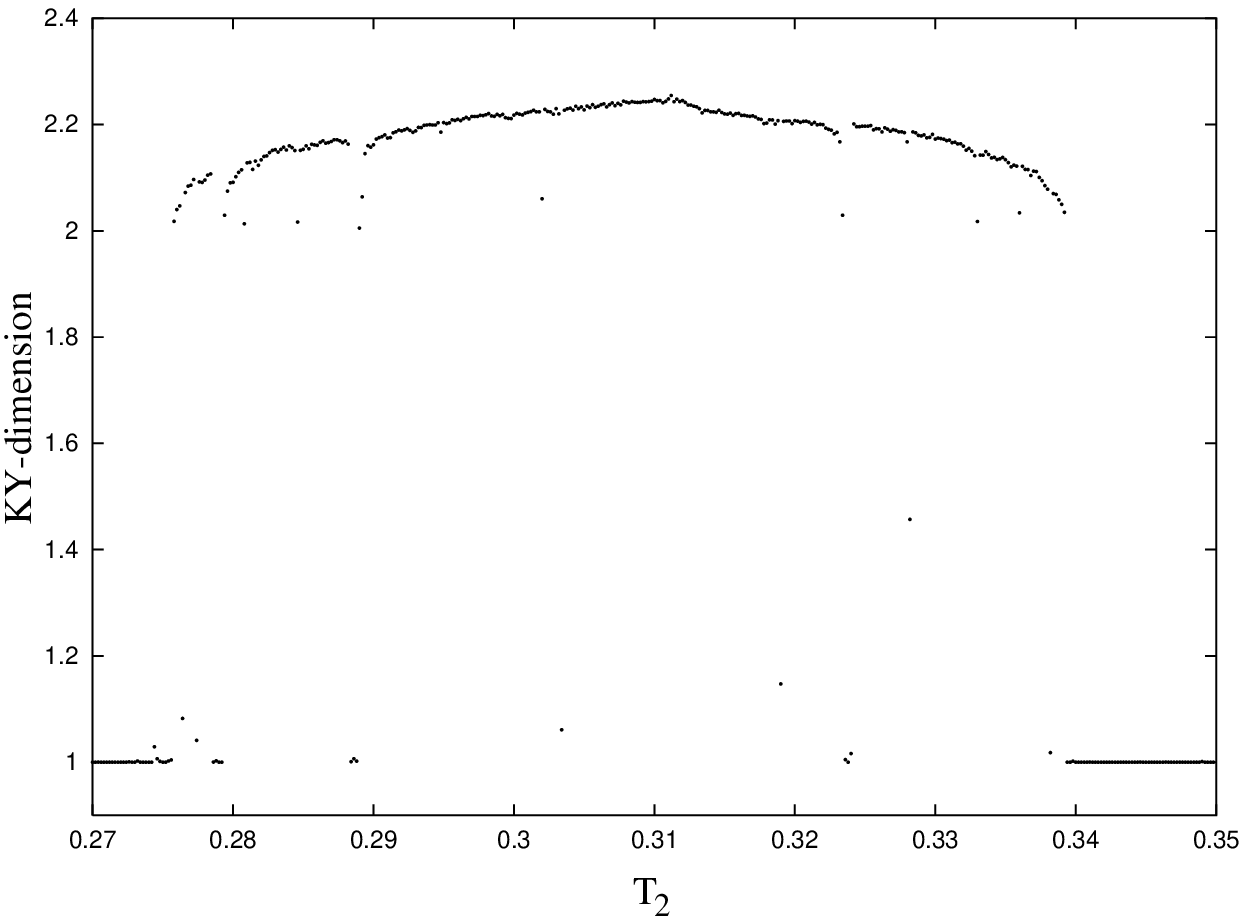,width=350pt}}
\end{picture}
\scaption{Numerical estimate of the Kaplan-Yorke dimension, obtained from an integration during 
$\Delta t=1.5\cdot 10^4$ at each parameter value $T_{2}$, for fixed $T_{1}=T^{(1)}$.}
\label{KYdim1}
\end{figure}

\subsection{Reduction to an invariant manifold}
\label{6to3}

With the parameters set to the values, specified in section \rf{6dbif}, we can scale the constants
in equations (\ref{odesys1}.1)-(\ref{odesys1}.3) as
\begin{alignat}{2}
\bar{C} &= \epsilon C & \qquad \kappa &= \epsilon \frac{\delta \lambda_{10}}{\lambda_{11}}=\epsilon \frac{\delta}{2}
\label{epsscaling}
\end{alignat}
where $\epsilon\approx 1/4$ and $\bar{C}\approx \kappa \approx 1$. System \rf{odesys1} is then written
symbolically as
\begin{align}
\label{symb}
\epsilon \dot{\mathbf{x}} &= \bar{\mathbf{f}}(\mathbf{x},\mathbf{y}) \tag{\ref{symb}.1} \\
\dot{\mathbf{y}} &= \mathbf{g}(\mathbf{x},\mathbf{y}) \tag{\ref{symb}.2}
\end{align}
\addtocounter{equation}{1}
where $\bar{\mathbf{f}}$ is defined as
\begin{align}
\label{fbar}
\bar{f}_{1} &= -\bar{C} (x_{1} - y_{1})  \tag{\ref{fbar}.1} \\
\bar{f}_{2} &= -\bar{C} (x_{2} - y_{2}) + \kappa (x_{1}x_{3}+ y_{1}y_{3}) \tag{\ref{fbar}.2} \\
\bar{f}_{3} &= -\bar{C} (x_{3} - y_{3}) - \kappa (x_{1}x_{2}+ y_{1}y_{2}) \tag{\ref{fbar}.3}
\end{align}
\addtocounter{equation}{1}
We assume, that there exists a globally attracting, three dimensional invariant manifold in system \rf{symb}, 
denoted by $W_{\epsilon}$.
This manifolds is represented as the graph of a function, $\phi_{\epsilon}$, of the baroclinic components:
\begin{equation}
W_{\epsilon} = \{ (\mathbf{x},\mathbf{y})\in \mathbb{R}^{6} | \mathbf{x}=\mathbf{\phi}_{\epsilon}(\mathbf{y}) \}
\label{W}
\end{equation}
and satisfies
\begin{equation}
\bar{\mathbf{f}}(\phi_{\epsilon},\mathbf{y})=\epsilon \mbox{D}\phi_{\epsilon}\cdot\mathbf{g}(\phi_{\epsilon},\mathbf{y})
\label{invariance}
\end{equation}
The solution of \rf{invariance} can be approximated asymptotically. Substituting the regular expansion \mbox{$\phi_{\epsilon}=\phi_{0}+\epsilon\phi_{1}+\ldots$}, we find
\begin{equation}
\mathbf{\phi}_{0}(\mathbf{y}) = \left( \begin{array}{ccc}
1 & 0 & 0 \\
0 & A(y_{1}) & B(y_{1}) \\
0 & -B(y_{1}) & A(y_{1})  \end{array} \right) \!\mathbf{y}
\label{order0}
\end{equation}
where $A$ and $B$ are defined as
\begin{alignat}{2}
A(y_{1}) &= \frac{\bar{C}^{2}-\kappa^{2}y_{1}^{2}}{\bar{C}^{2}+\kappa^{2}y_{1}^{2}}=\cos \gamma & \quad \text{and}\quad
B(y_{1}) &= \frac{2\bar{C}\kappa y_{1}}{\bar{C}^{2}+\kappa^{2}y_{1}^{2}}=\sin \gamma.
\label{ABdef}
\end{alignat}
This zeroeth order approximation  has a clear physical interpretation. 
It describes a zonally symmetric part of the flow which
is equivalent barotropic (i.e. $\Psi_{3}\propto\Psi_{1}$) and a phase shift, $\gamma$, between the traveling waves in
the upper and the lower layer. 

The nonlinear terms in equations \rf{odesys1} can be divided into two groups:
one represents advection of waves with the zonally symmetric flow (the terms proportional to $x_{1}$) and
the other represents energy exchange between waves and the zonally symmetric flow. The nonlinear terms in
equation (\ref{odesys1}.4) belong to the second group. If the phase shift is zero, these terms cancel, and
the wave components are decoupled from the zonally symmetric components.
Thus, the function $\phi_{0}$ describes how the energy transfer depends on strength of the
zonally symmetric flow.

The asymptotic expansion of $\phi_{\epsilon}$ can not accurately describe the solution of equation \rf{invariance}
at the realistic parameter value $\epsilon=0.25$. Therefore, we use a numerical algorithm to continue
the solution, known analytically in the limit of $\epsilon\downarrow 0$. The algorithm is similar to the 
graph transform described in \cite{bro1}. In contrast to the systems considered in their work, however,
ours is a continuous time system. Therefore, we apply 
the graph transform to the map, induced by an implicit Euler 
step which approximates the flow of system \rf{symb} over a finite time interval.
This approach was already used by \cite{foia} to approximate inertial manifolds by their
modified Galerkin method.

Another difference is the choice of coordinates. Here, we do not represent the (approximate) invariant manifold 
as the graph of a section of the normal
bundle of some given approximation. Instead, we globally represent it as the graph of
a function $\phi_{\epsilon}(\mathbf{y})$. This can be done provided that $\mbox{D}\phi_{\epsilon}$
has full rank. For small $\epsilon$ this condition is satisfied, as we have 
$\,\det\mbox{D}\phi_{\epsilon}=1+\mbox{O}(\epsilon)$. When increasing $\epsilon$ the condition has to be
checked numerically.

Suppose that, for some fixed value of $\epsilon$, we have an approximation, $W$, of $W_{\epsilon}$.
By assumption, $W_{\epsilon}$ is globally attracting so that the image of $W$ under the flow over
a finite time interval of system \rf{symb} lies closer to $W_{\epsilon}$ than $W$.
Thus, by approximating the flow of system \rf{symb}, we can calculate an improved approximation $\bar{W}$.
For this end we use the map $E: \mathbb{R}^{6}\rightarrow\mathbb{R}^{6}$, where 
$(\bar{\mathbf{x}},\bar{\mathbf{y}})=E((\mathbf{x},\mathbf{y}))$ is the solution of
\begin{align}
\label{imeul}
\bar{\mathbf{x}} &= \mathbf{x} +\frac{\Delta}{\epsilon} \bar{\mathbf{f}}(\bar{\mathbf{x}},\bar{\mathbf{y}}) \tag{\ref{imeul}.1} \\
\bar{\mathbf{y}} &= \mathbf{y} +\Delta \mathbf{g}(\bar{\mathbf{x}},\bar{\mathbf{y}}) \tag{\ref{imeul}.2}
\end{align}\addtocounter{equation}{1}which defines an implicit Euler time step. The step size, $\Delta$, is a free parameter.
The improved approximation is then defined as $\bar{W}=E(W)$.

In order to represent $\bar{W}$ as the graph of a function $\bar{\phi}(\mathbf{y})$, we map
a point $(\mathbf{\phi}(\mathbf{y}),\mathbf{y})\in W$ onto the point $(\bar{\mathbf{\phi}}(\mathbf{y}),\mathbf{y})\in\bar{W}$,
where $\bar{\phi}$ is the solution of
\begin{equation}
\bar{\mathbf{\phi}}=\phi(\mathbf{y}-\Delta \mathbf{g}(\bar{\mathbf{\phi}},\mathbf{y}))+\frac{\Delta}{\epsilon}\bar{\mathbf{f}}(\bar{\mathbf{\phi}},\mathbf{y}) \label{NRE}
\end{equation}
In other words, for a given vector $\mathbf{y}$ we look for the point $(\mathbf{\phi}(\mathbf{y}'),\mathbf{y}')\in W$
which is mapped according to \rf{imeul} onto  $(\bar{\mathbf{\phi}}(\mathbf{y}),\mathbf{y})\in\bar{W}$.
In figure \rf{schematic} this procedure is sketched.
\begin{figure}[t]
\begin{picture}(350,250)
\centerline{\epsfig{file=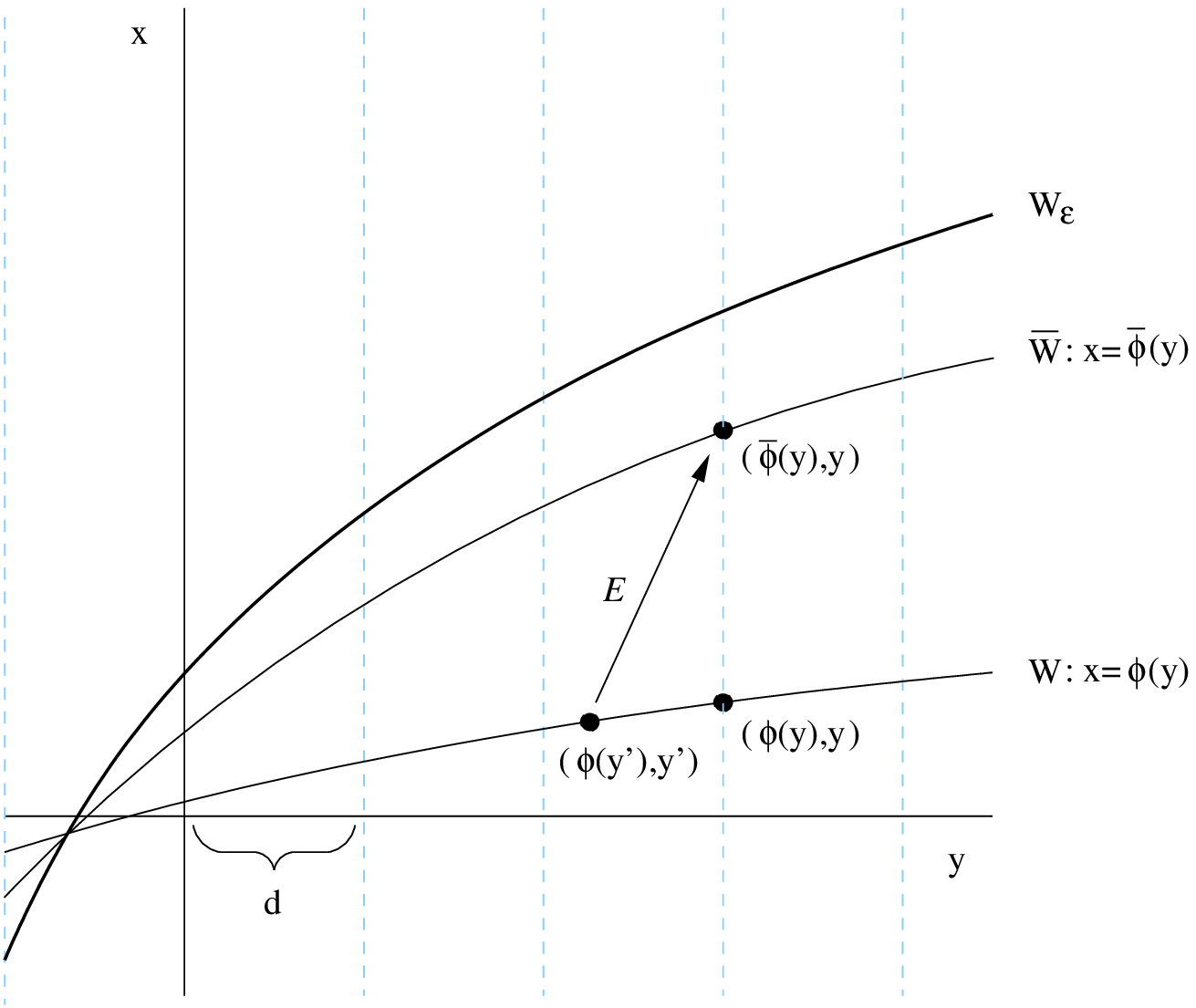,height=250pt}}
\end{picture}
\scaption{Schematic picture of the graph transform. Note, that equilibria of system \rf{symb} are intersection
points of $W$, $\bar{W}$ and $W_{\epsilon}$. The dashed lines represent the mesh $\mathbf{y}_{ijk}$.}
\label{schematic}
\end{figure}

Differentiating equation \rf{NRE} with respect to $\mathbf{y}$, we find
\begin{equation}
(\mathbb{I}-\frac{\Delta}{\epsilon}\mbox{D}_{x}\bar{\mathbf{f}}
+\Delta\mbox{D}\mathbf{\phi}|_{\mathbf{y}'} \, \mbox{D}_{x}\mathbf{g})\, \mbox{D}_{y}\bar{\mathbf{\phi}}
=\mbox{D}\mathbf{\phi}|_{\mathbf{y}'} \, (\mathbb{I}-\Delta\mbox{D}_{y}\mathbf{g})
+\frac{\Delta}{\epsilon}\mbox{D}_{y}\bar{\mathbf{f}} \label{tangent}
\end{equation}
from which we can calculate $\mbox{D}_{y}\bar{\mathbf{\phi}}$, and thus the tangent space $T_{(\bar{\mathbf{\phi}},\mathbf{y})}\bar{W}$, from $\mbox{D}_{y}\mathbf{\phi}$ and the Jacobian of system \rf{symb}. 
This enables us to define a local error function. Let $F=(\bar{\mathbf{f}}/\epsilon,\mathbf{g})$
denote the vector field \rf{symb} and $\mathcal{P}$ the orthogonal projection onto 
the tangent space at $(\bar{\mathbf{\phi}},\mathbf{y})$. Then we define the error function
\begin{equation}
e(\mathbf{y})= \frac{2}{\pi}\arccos \frac{\| \mathcal{P}F(\bar{\mathbf{\phi}},\mathbf{y})\| }{\| F(\bar{\mathbf{\phi}},\mathbf{y})\| }
\end{equation}

We will represent the approximate invariant manifold on a cubic lattice in $\mathbb{R}^3$. The vertices of this
lattice are located at $\mathbf{y}_{ijk}=d\, (i,j,k)^T$, where d is the lattice spacing and $i,j,k$ are integers.
We consider a finite number of lattice points, demanding that all points $(\phi_{0}(\mathbf{y}_{ijk}),\mathbf{y}_{ijk})$
lie within the trapping region defined in section \rf{sixD}. In fact, inequality \rf{trap} is a fairly course estimate.
In order to reduce computation time and data storage, we use a sharper estimate, obtained by numerical computation
of the eigenvalues of the linear part of equations (\ref{odesys1}.1-\ref{odesys1}.6). To an approximate solution 
$\bar{\mathbf{\phi}}$ we can
then assign the error
\begin{equation}
\mathcal{E}=\max_{ijk} \, e(\mathbf{y}_{ijk})
\end{equation}

In order to approximate the invariant manifold for finite $\epsilon$, we will proceed in steps.
First of all, we fix an initial value for $\epsilon$ and calculate 
$\phi_{1}$ and $\mbox{D}\phi_{1}$ on each lattice point. 
Next, we solve equation \rf{NRE}, by Newton iteration, and subsequently \rf{tangent}, 
to find the next approximation $\bar{\mathbf{\phi}}$ and
its derivatives. 
A suitable initial guess for the Newton iteration is obtained by linearisation in 
$\mathbf{\xi}=\mathbf{y}'-\mathbf{y}$. This yields
\begin{equation}
\left( \begin{array}{c}
\!\! \bar{\mathbf{\phi}}(\mathbf{y}) \!\! \\
\!\! \mathbf{y} \!\! \end{array}\right)
=E \left( \begin{array}{c}
\!\! \mathbf{\phi}(\mathbf{y}') \!\! \\
\!\! \mathbf{y}' \!\! \end{array}\right)
\approx E \left( \begin{array}{c}
\!\! \mathbf{\phi}(\mathbf{y}) \!\! \\
\!\! \mathbf{y} \!\! \end{array}\right) +\mbox{D}E \cdot
\left( \begin{array}{c}
\!\! \mbox{D}\mathbf{\phi} \!\! \\
\!\! \mathbb{I} \!\! \end{array} \right)
\cdot \mathbf{\xi}
\end{equation}
from which we can estimate $\mathbf{\xi}$ and, subsequently, $\bar{\phi}$.
The graph transform is iterated untill the global error is smaller
than a fixed threshold $\mathcal{E}_{max}$. If this is achieved, we increase epsilon and iterate the process.

When solving equation \rf{NRE}, evaluation of $\phi$ and $\mbox{D}\phi$ inbetween lattice points is necessary. This is
done by linear interpolation. To this end, each cube in the lattice is divided into six tetraeders of equal
volume. The associated error is expected to be of order $\mbox{O}(d^{2})$. At the edge of the domain we
consider, it may happen, that evaluation of $\phi$ outside this domain is required. If so, we solve equation
\rf{NRE}, substituting $\phi (\mathbf{y}') \rightarrow \phi (\mathbf{y})+\mbox{D}\phi \cdot \xi$. The derivatives
of $\bar{\mathbf{\phi}}$ are then calculated by finite difference. 
This, however, introduces an error, which disables us to 
continue the invariant manifold up to $\epsilon=0.25$.

For example we approximate the manifold at parameter values $(T_{1},T_{2})=(0.6 , 0.25)$. The step size is
fixed to $\Delta= 0.0025$ and the increment of $\epsilon$ is chosen in the range $[0.001,0.005]$.
The maximal error is fixed to $\mathcal{E}_{max}=0.05$. 
At each value of $\epsilon$ about six iterations of the graph transform
are needed. The computations were done with a lattice spacing of $d=0.01$, in the trapping region
$L\leq 0.52$. Thus, about $5\cdot 10^{5}$ points on the manifold are calculated. 

In figure \rf{ilustr} the approximate invariant manifold is illustrated. Shown are the stable periodic
orbit of system \rf{symb} at $\epsilon=0.1$ and $(T_{1},T_{2})=(0.6 , 0.25)$ and a forward integration
of the system $\dot{\mathbf{y}}=\mathbf{g}(\phi_{\epsilon}(\mathbf{y}),\mathbf{y})$. To find the latter
integral curve we approximated every point of $\phi_{\epsilon}(\mathbf{y})$
needed by linear interpolation of the results of the calculation presented above. 
\begin{figure}[h]
\begin{picture}(350,170)
\centerline{\epsfig{file=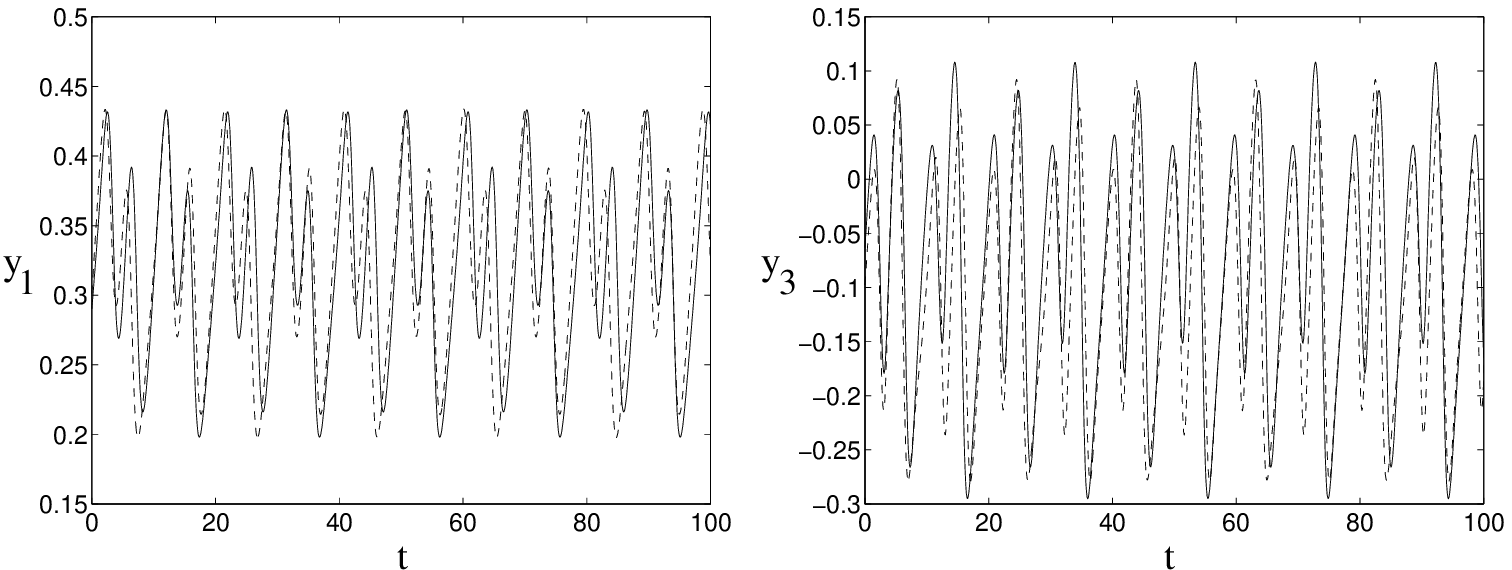,width=430pt}}
\end{picture}
\scaption{Time series of the zonally symmetric component, $y_{1}$, and a wave component, $y_{3}$. Solid:
solution of system \rf{symb} at $(T_{1},T_{2})=(0.6,0.25)$ and $\epsilon=0.1$. Dashed: solution of
$\dot{\mathbf{y}}=\mathbf{g}(\phi_{\epsilon}(\mathbf{y}),\mathbf{y})$, using the approximate solution 
$\phi_{\epsilon}$ of equation \rf{invariance} obtained by the graph transform method.}
\label{ilustr}
\end{figure}

\section{The reduced system}

In the following, we will describe system \rf{symb}, reduced to the approximate invariant manifold $W_{0}$.
This reduction can be done analytically, so that we can compare bifurcation diagrams. Physically, an argument
to study the system reduced to $W_{0}$, rather than to the numerically approximated $W_{\epsilon}$, is that
the quantitative error, introduced by setting $\epsilon=0$, is smaller than the error introduced by the Galerkin
approximation. Equations \rf{odesys1} form a qualitative model of one aspect of the atmospheric circulation,
namely the interaction between the jet stream and the baroclinic waves. A further simplification is justified
if it keeps the qualitative behaviour in tact.

Substituting expression \rf{order0}, we find that the reduced system, $\dot{\mathbf{y}}=\mathbf{g}(\phi_{0}(\mathbf{y})
,\mathbf{y})$, is given by
\begin{align}\label{odesys2}
\dot{y}_{1} &= -c_{1}y_{1}-d_{1} (y_{2}^2 +y_{3}^2) +\bar{T}_{1} \tag{\ref{odesys2}.1} \\
\dot{y}_{2} &= -c_{2}y_{2}+c_{3} y_{3}+d_{3} y_{1}y_{3}+d_{2} y_{1}y_{2} +\bar{T}_{2} \tag{\ref{odesys2}.2} \\
\dot{y}_{3} &= -c_{2}y_{3}-c_{3} y_{2}-d_{3} y_{1}y_{2}+d_{2} y_{1}y_{3} +\bar{T}_{3} \tag{\ref{odesys2}.3}
\end{align}\addtocounter{equation}{1}
where we have introduced
\begin{alignat}{3}
c_{1} &= (-2\alpha\lambda_{01}C'+h_{N})/\bar{\lambda}_{01} & \qquad \bar{T}_{1} &= h_{N}T_{1}/\bar{\lambda}_{01}
& \qquad d_{1} &= \delta B/\bar{\lambda}_{01} \nonumber \\
c_{2} &= (-\alpha\lambda_{11}[C(1-A)+2C']+h_{N})/\bar{\lambda}_{11} & \qquad \bar{T}_{2} &= h_{N}T_{2}/\bar{\lambda}_{11} & \qquad d_{2} &= \delta\nu B/\bar{\lambda}_{11} \nonumber \\
c_{3} &= -\alpha\lambda_{11}CB/\bar{\lambda}_{11} & \qquad \bar{T}_{3} &= h_{N}T_{3}/\bar{\lambda}_{11} & \qquad
d_{3} &= \delta[\bar{\lambda}_{10}-\nu_{10}A]/\bar{\lambda}_{11} \label{defs2}
\end{alignat}
Again, we have a Lyapunov function
\begin{equation}
L=y_{1}^2 +\frac{\bar{\lambda}_{11}}{\nu_{10}\bar{\lambda}_{01}}(y_{2}^2 +y_{3}^2)
\label{lyapunov2}
\end{equation}
and a trapping region defined by
\begin{equation}
L \leq \left( \frac{\bar{\lambda}_{11}}{\nu_{10}\bar{\lambda}_{01}} \right) ^{2} \frac{\| \bar{T}\| ^{2}}{c_{1}^2}
\label{trap2}
\end{equation}
In contrast to equations \rf{odesys1}, the divergence of the vector field of this reduced system can change
sign, so that volume elements are not necessarily shrinking.

As explained in section \rf{6to3}, the amplitude of the nonlinear interaction depends on $y_{1}$ through
$A$ and $B$. In figure \rf{B} the dependence of the interaction coefficient in (\ref{odesys2}.1) is shown.
\begin{figure}[t]
\begin{picture}(350,250)
\centerline{\epsfig{file=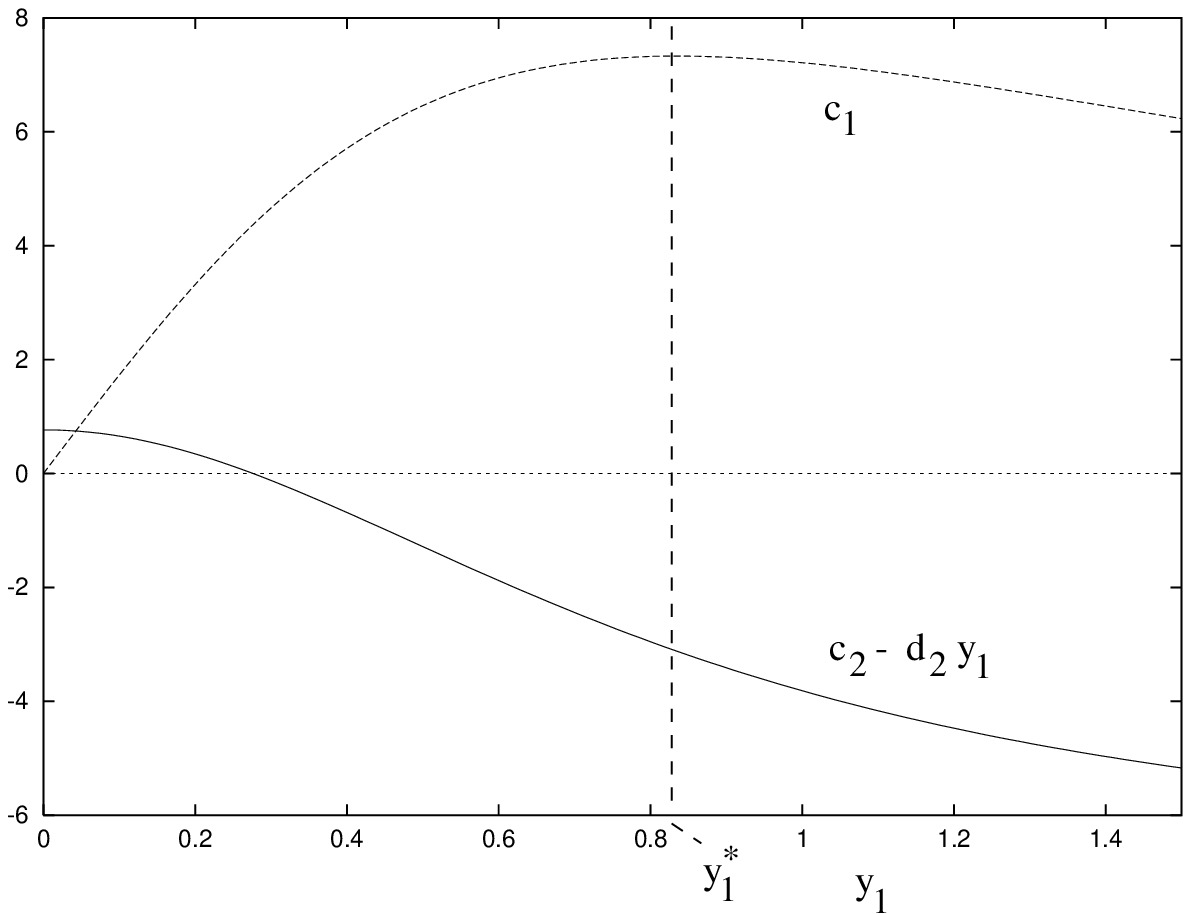,height=250pt}}
\end{picture}
\scaption{Dashed: the coefficient of interaction, $d_{1}$, between the jet stream pattern and the wave patterns. Solid:
the effective damping coefficient of $y_{2,3}$. 
If we fix $y_{1}=y_{1}^{*}=\bar{C}/\kappa$ model \rf{odesys2} is equivalent to the Lorenz-84 model described in section \rf{Lor84}}
\label{B}
\end{figure}
This figure illustrates the life cycle of the baroclinic waves. The linear stability of the Hadley circulation
is determined by the effective damping coefficient of the wave components $y_{2,3}$, given by $c_{2}-d_{3}y_{1}$.
If this number is positive, the waves are damped. If it is negative, a perturbation of the Hadley circulation will
lead to growing waves. 
If $y_{1}$ is small, there is little energy transfer and the waves are damped. 
As $y_{1}$ is forced by the meridional temperature gradient,
$\bar{T}_{1}$, it grows and the energy transfer increases. Beyond $y_{1}\approx 0.276$, the effective damping 
coefficient is negative, and baroclinic waves can grow.
When $y_{1}=\bar{C}/\kappa \approx 0.83$, the phase shift, 
$\gamma$, is maximal and the waves are optimally baroclinic. They extract energy from the jet stream and $y_{1}$
decreases. Then the waves decay and become decreasingly baroclinic in the process.

In order to see, if system \rf{odesys2} behaves qualitatively the same as system \rf{odesys1}, we study the 
bifurcation diagram. It is shown in figure \rf{diagram2}.
\begin{figure}[h!]
\begin{picture}(400,330)
\centerline{\epsfig{file=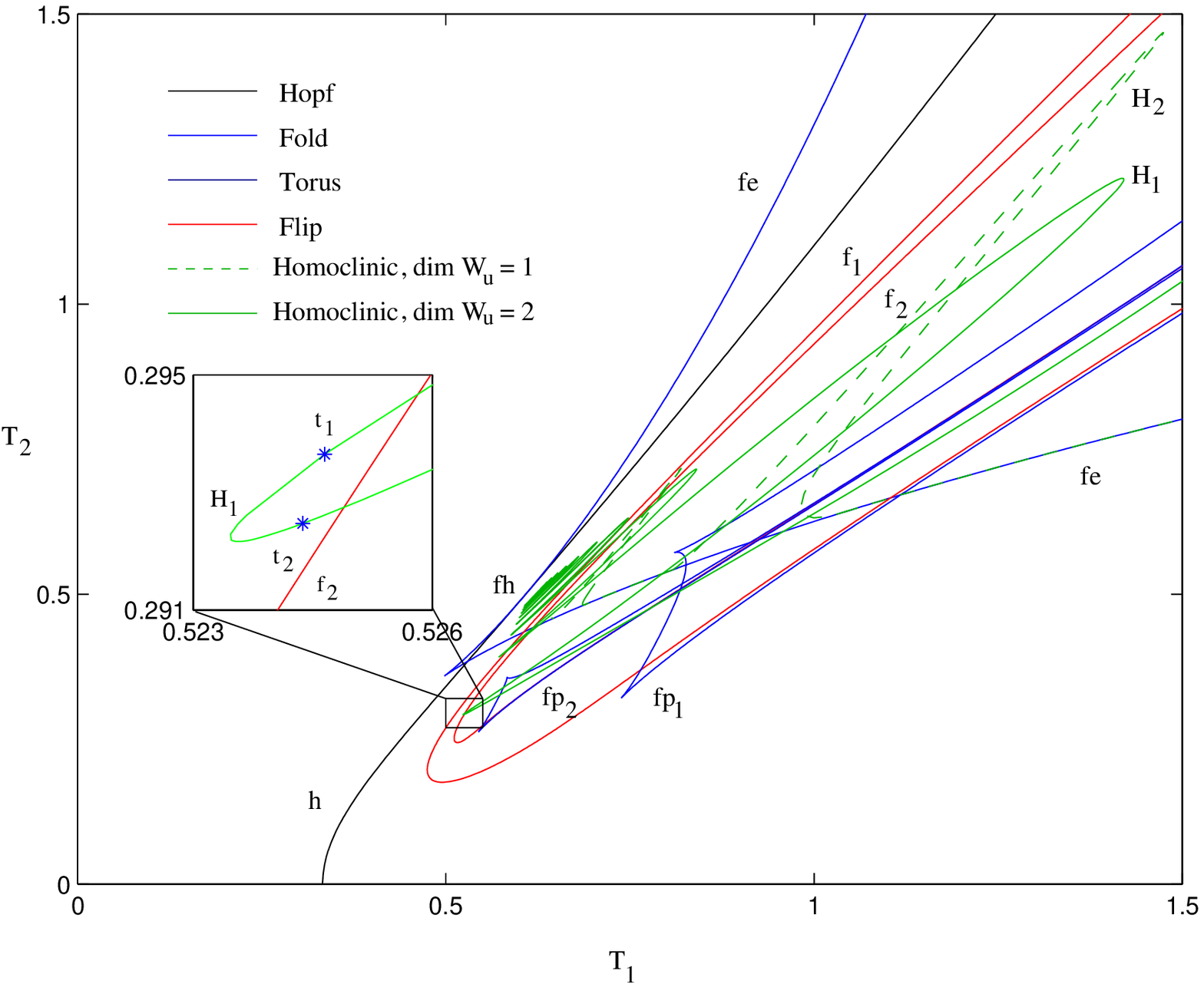,width=400pt}}
\end{picture}
\scaption{Bifurcation diagram of system \rf{odesys2}. The enlargement shows the neutral saddle focus
transitions $\mbox{t}_{1,2}$. Note the similarity to diagram \rf{diagram1}}
\label{diagram2}
\end{figure}
All the bifurcations displayed in diagram \rf{diagram1} are present here, too. Therefore, the qualitative behaviour
of the reduced system \rf{odesys2} is the same as that of the full system \rf{odesys1}. The essence of the extra
degrees of freedom in the six dimensional model can be captured by the variable coefficients in the three dimensional
model.

\section{Reduction to the Lorenz-84 model}
\label{Lor84}
\begin{figure}[ht]
\begin{picture}(400,330)
\centerline{\epsfig{file=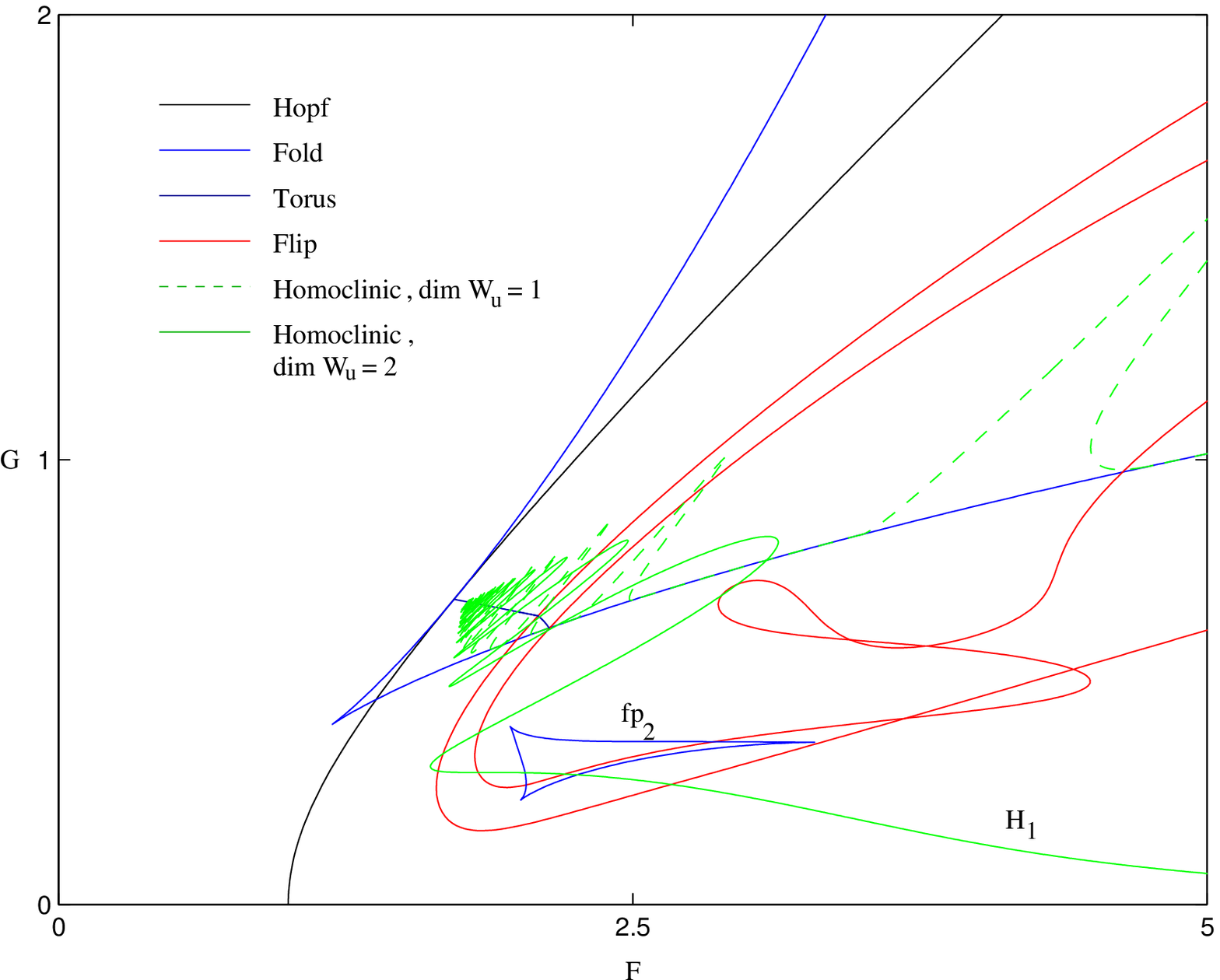,width=400pt}}
\end{picture}
\scaption{Bifurcation diagram of system \rf{odesys3}, with parameters $a$ and $b$ as calculated from the physical
parameters $C$, $C'$, $h_{N}$ and $\alpha$.}
\label{diagram3}
\end{figure}
As a final simplification of our model, we fix the coefficients in equations \rf{odesys2}. The choice 
\begin{alignat}{2}
A=A(\bar{C}/\kappa)&=0 & \qquad B=B(\bar{C}/\kappa) &= 1 \nonumber
\end{alignat}
maximizes the efficiency of the energy transfer between the baroclinic waves and the jet stream.
The phase shift is fixed to $\gamma=\pi/2$. Equivalently, we can reduce model \rf{odesys1} to the
tangent space $\mbox{T}_{(\mathbf{x}^{*},\mathbf{y}^{*})}W_{0}$, where 
$(\mathbf{x}^{*},\mathbf{y}^{*})=(\bar{C}/\kappa,0,0,\bar{C}/\kappa,0,0)$. In other words, we linearise
$\phi_{0}$ around the Hadley state.

We then scale $\mathbf{y}$ and $t$ according to
\begin{alignat}{2}
t &= [c_{2}+c_{3}\frac{d_{2}}{d_{3}}]^{-1} t' & \qquad y_{1} &= \frac{1}{d_{2}}[c_{2}+c_{3}\frac{d_{2}}{d_{3}}]x-\frac{c_{3}}{d_{3}} \nonumber \\
y_{2} &= [c_{2}+c_{3}\frac{d_{2}}{d_{3}}]\frac{1}{\sqrt{d_{1}d_{2}}} \,y & \qquad y_{3} &=[c_{2}+c_{3}\frac{d_{2}}{d_{3}}]\frac{1}{\sqrt{d_{1}d_{2}}}  \,z \label{transform1}
\end{alignat}
and find
\begin{align}\label{odesys3}
\dot{x} & = -y^{2}-z^{2}-a x + a F 		\tag{\ref{odesys3}.1} \\
\dot{y} & = x y- b x z - y + G		\tag{\ref{odesys3}.2} \\
\dot{z} & = b x y + x z - z			\tag{\ref{odesys3}.2}
\end{align}\addtocounter{equation}{1}
which is the model introduced by \cite{lor2}. For the parameters we find
\begin{alignat}{2}
F &= d_{2}[c_{2}+c_{3}\frac{d_{2}}{d_{3}}]^{-2}(\bar{T}_{1}+\frac{c_{1}c_{3}}{d_{3}}) & \qquad G &= \sqrt{d_{1}d_{2}}
[c_{2}+c_{3}\frac{d_{2}}{d_{3}}]^{-2} \bar{T}_{2} \nonumber \\
a &= c_{1}[c_{2}+c_{3}\frac{d_{2}}{d_{3}}]^{-1} & \qquad b &= \frac{d_{3}}{d_{2}} \label{transform2}
\end{alignat}
With the parameters as specified in section \rf{6dbif}, we thus obtain
\begin{alignat}{2}
a &\approx 0.35 & \qquad b &\approx 1.33 \label{newvals}
\end{alignat}
in contrast to the traditional values $a=1/4$ and $b=4$. Diagram \rf{diagram3} has been obtained by setting
the parameters according to \rf{newvals}. There still is a strong similarity to diagrams \rf{diagram1} and \rf{diagram2}.
However, the neutral saddle focus transitions have disappeared. Along both homoclinic bifurcation curves
only a stable cycle is created. The accumulation of $1\!:\!2$ resonance points is still there and,
although homoclinic curve $\mbox{H}_{1}$ moves farther away, the
branch switching mechanism along $\mbox{fp}_{2}$ works as described in section \rf{6dbif}.

\subsection{Continuation in $a$ and $b$}

Finally, we continue the bifurcations in diagram \rf{diagram3} in parameters $a$ and $b$ in order to
establish the relation to the diagram at traditional parameter values, presented in \cite{shil}.
\begin{figure}[t]
\begin{picture}(400,330)
\centerline{\epsfig{file=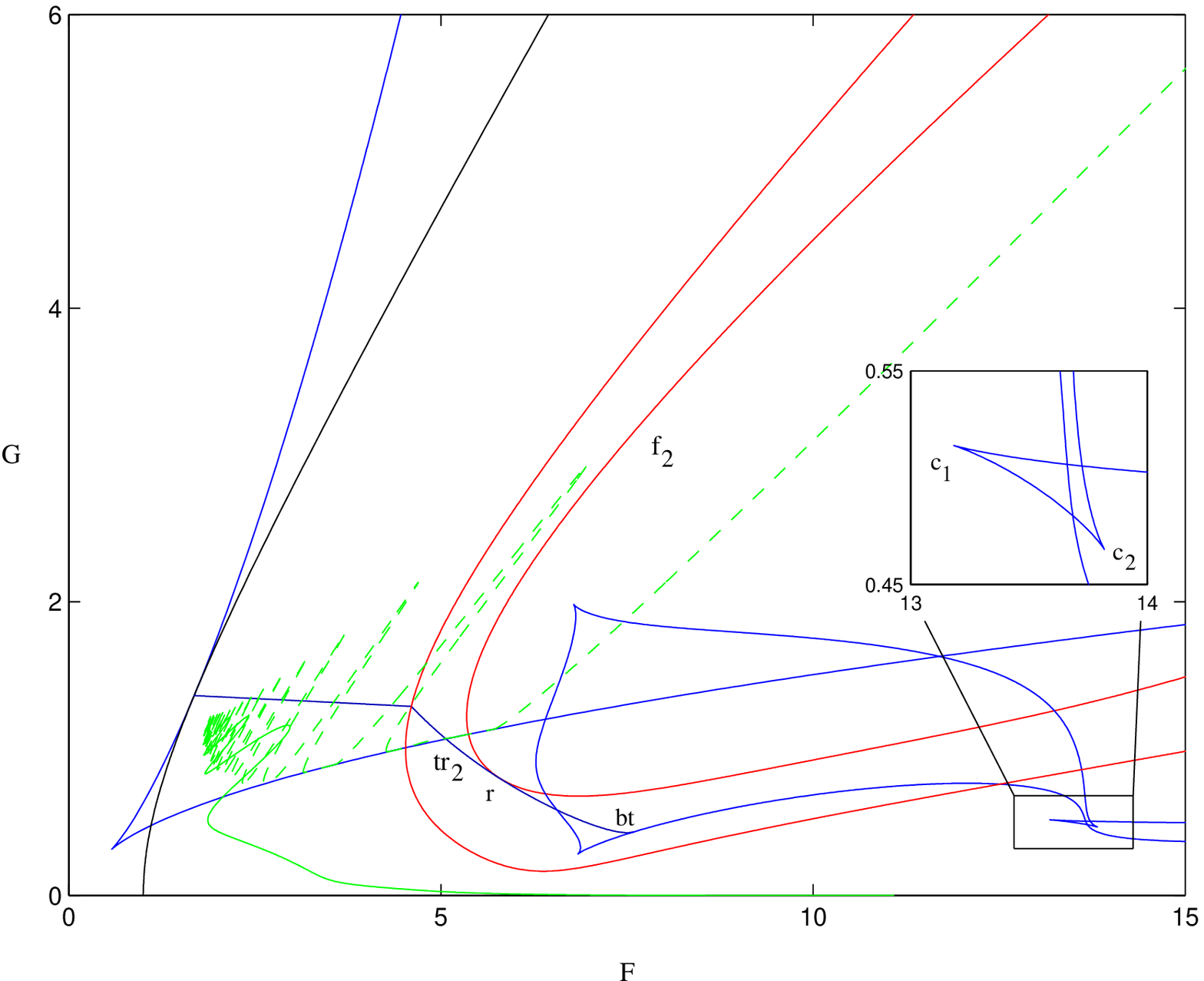,width=400pt}}
\end{picture}
\scaption{Bifurcation diagram of system \rf{odesys3}, with $a=1/4$ and $b=3.24$. Torus bifurcation line $\mbox{tr}_{2}$
and flip bifurcation line $\mbox{f}_{2}$ are tangent at point r. $\mbox{tr}_{2}$ ends in a $1\!:\!1$ resonance, or
Bogdanov-Takens point, marked bt. The enlargement shows that extra cusps, $\mbox{c}_{1,2}$, have developed
on saddle node line $\mbox{fp}_{2}$. Colour coding as in diagram \rf{diagram3}.}
\label{diagram4}
\end{figure}
Changing $a$ to its traditional value, $a=1/4$, does not change the bifurcation diagram qualitatively.
When changing $b$ two changes are apparent. Shown in figure \rf{diagram4} is the diagram obtained for
$a=1/4$ and $b=3.24$. The second torus bifurcation line, $\mbox{tr}_{2}$, connecting two $1\!:\!2$ resonance
points in diagram \rf{diagram3}, is now tangent to flip bifurcation line $\mbox{f}_{2}$ at point $\mbox{r}$
and connects
to the fold line $\mbox{fp}_{2}$. These two meet in a $1\!:\!1$ resonance points, also called 
Bogdanov-Takens point.

Also, a pair of cusp points, marked $\mbox{c}_{1,2}$, has developed on fold line $\mbox{fp}_{2}$. These
cusps denote creation and vanishing of successive wiggles on the branch of periodic solutions connected
to homoclinic curve $\mbox{H}_{1}$.

\begin{figure}[ht]
\begin{picture}(400,330)
\centerline{\epsfig{file=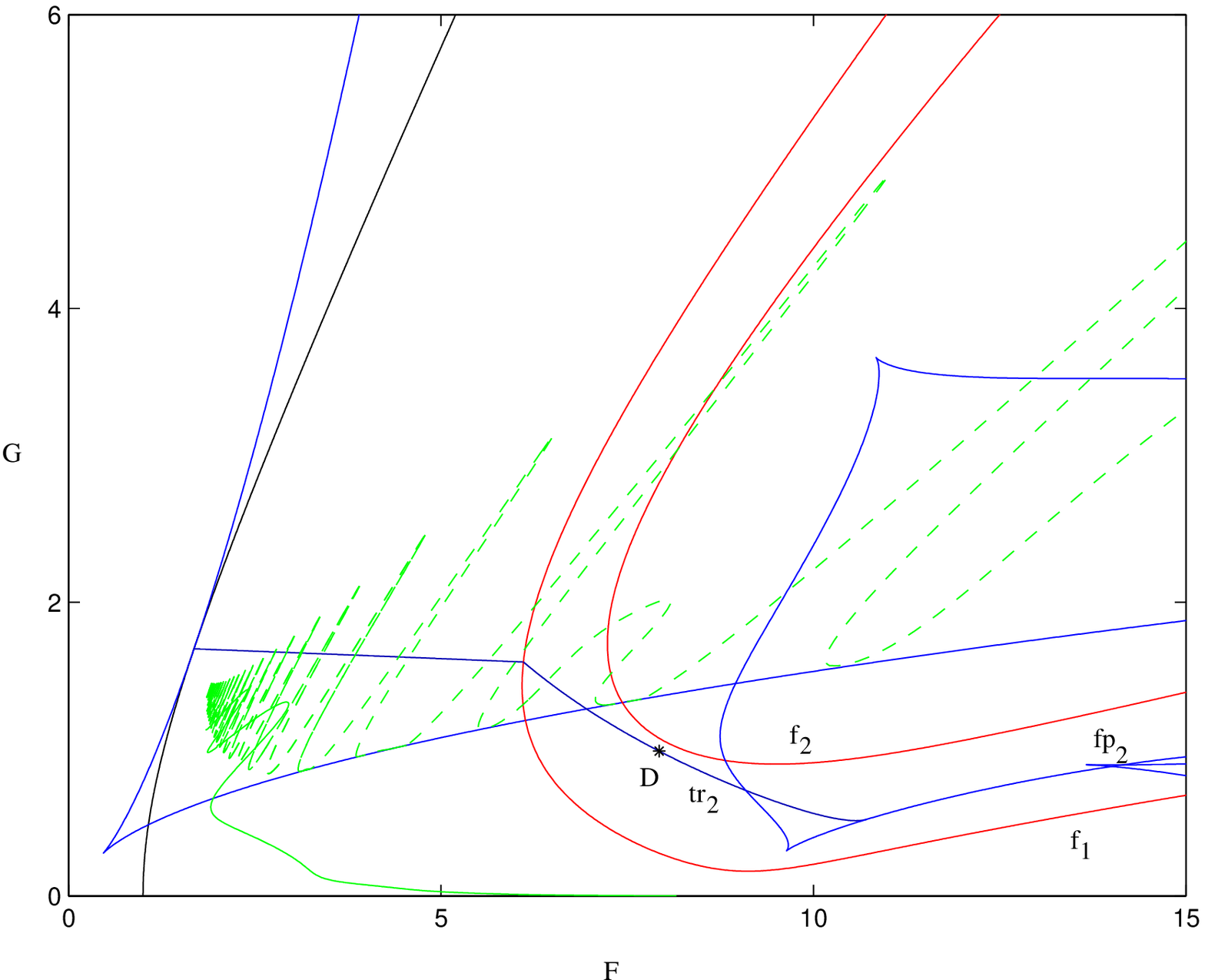,width=400pt}}
\end{picture}
\scaption{Bifurcation diagram of system \rf{odesys3}, with $a=1/4$ and $b=4$, see also \cite{shil}. Colour
coding as in diagram \rf{diagram3}. The angular degeneracy point on $\mbox{tr}_{2}$ is marked $\mbox{D}$.}
\label{diagram5}
\end{figure}
Figure \rf{diagram5} shows the bifurcation diagram at parameter values $a=1/4$ and $b=4$. This diagram
was presented by \cite{shil}. The torus bifurcation line $\mbox{tr}_{2}$ no longer connects to flip 
bifurcation line $\mbox{f}_{2}$ and, consequently, has developed an angular degeneracy (\cite{peck})
at point $\mbox{D}$. Along curve $\mbox{tr}_{2}$ the multipliers of the periodic orbit are given
by $\exp(\pm\smi \phi)$, where $\phi$ is the phase angle. At point $\mbox{D}$ it has a maximum
given by $\phi\approx 0.8 \pi$.

If the forcing is set to $(F,G)=(8,1)$ the model is known to behave
chaotically (e.g. \cite{lor2}). An important, unanswered question is how this chaotic behaviour is brought
about. With the derivation of the model and the bifurcation analysis in mind, we discuss four
different possible routes to chaos in the following subsections.

\subsubsection{Period doubling cascades}

The flip bifurcation lines $f_{1,2}$ in diagram \rf{diagram5} still are the first of an infinite
sequence. However, as explained in section \rf{6dbif}, figure \rf{branchsw}, a cross section with
$F$ fixed does not show a period doubling cascade followed by an inverse cascade, due
to branch switching along the cusped saddle node line $\mbox{fp}_{2}$. The flip bifurcation
lines $\mbox{f}_{1,2,\ldots}$, extending beyond the limits of diagrams \rf{diagram3}-\rf{diagram5},
are closed curves. As can be seen in diagram \rf{diagram3}, these curves form unnested islands.
In accordance with the claim to general applicability of the analysis presented in \cite{wiec},
the bifurcation diagram of the Lorenz-84 model has much in common with that of their rate equation
model. Chaotic attractors can be created and destroyed through different routes, including
period doubling cascades, the Ruelle-Takens scenario and intermittency.

\subsubsection{Ruelle-Takens scenario}
\begin{figure}[t]
\begin{picture}(400,330)
\centerline{\epsfig{file=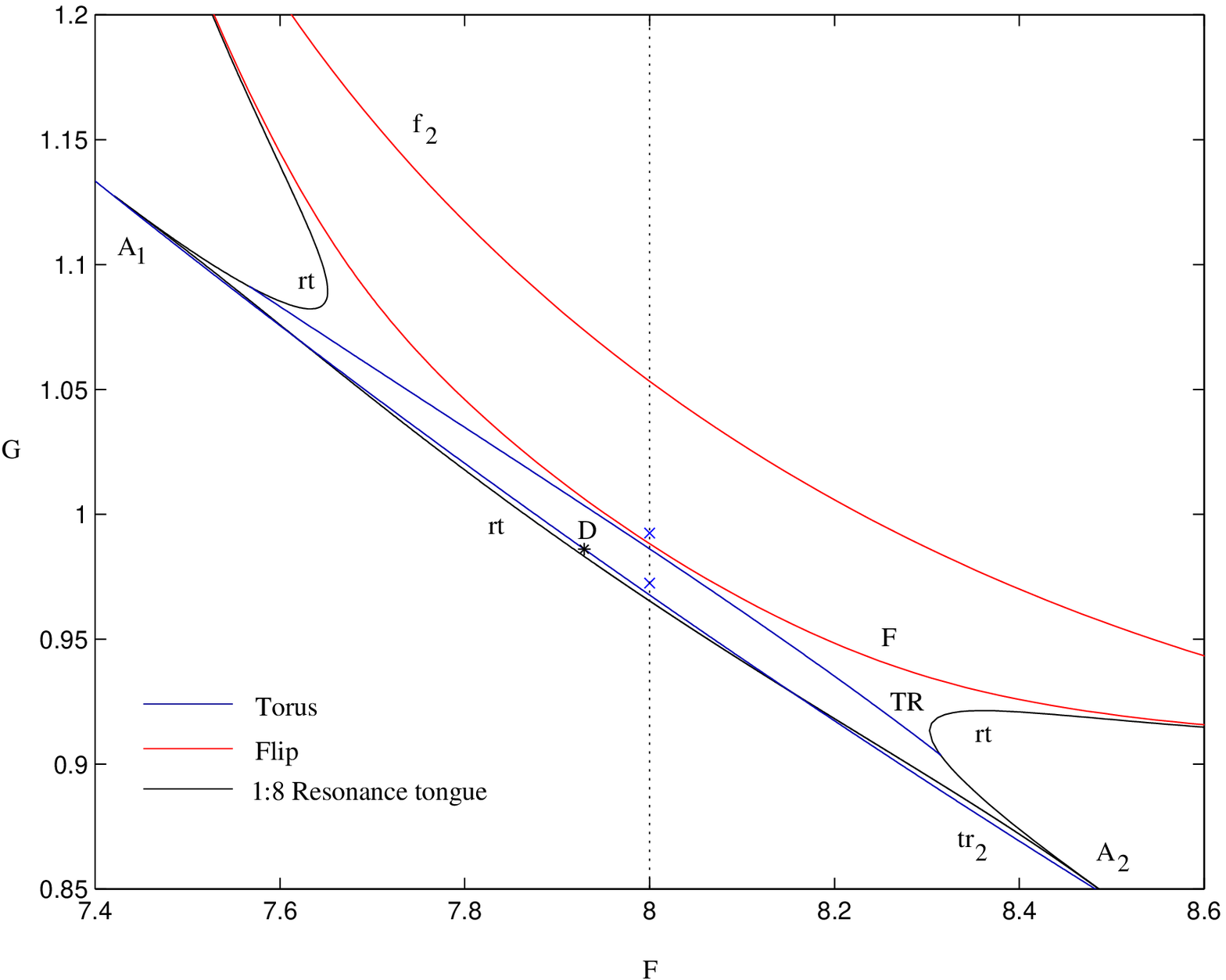,width=400pt}}
\end{picture}
\scaption{Detail of diagram \rf{diagram5}. Added: boundaries of the $8\!:\!1$ resonance tongue
and a torus and flip bifurcation of the period 8 orbit. The crosses mark the parameter values
of Poincar\'e sections in figure \rf{secs}.}
\label{resonance}
\end{figure}
Another possible route to chaos was proposed by \cite{ruel}. In this scenario, an invariant torus
is created first. Then a periodic obit appears on the torus, when crossing the boundary of an
Arnold resonance tongue. If this periodic orbit bifurcates, a chaotic attractor can be created.
This scenario can be observed in the Lorenz-84 model. We will concentrate on the chaotic behaviour
at parameter values $(F,G)=(8,1)$, close to the angular degeneracy point $\mbox{D}$.

If we fix $F=8$ and increase $G$ from below $\mbox{tr}_{2}$, 
we find the attracting period two orbit first, then quasiperiodic behaviour in a small interval,
then a period 8 orbit on the invariant torus. Figure \rf{resonance} shows a detail of
diagram \rf{diagram5}, with the boundaries of the $8\!:\!1$ resonance tongue and two
subsequent bifurcations, a torus and a flip bifurcation, of the period 8 orbit. 

\begin{figure}[t]
\begin{picture}(400,150)
\centerline{\epsfig{file=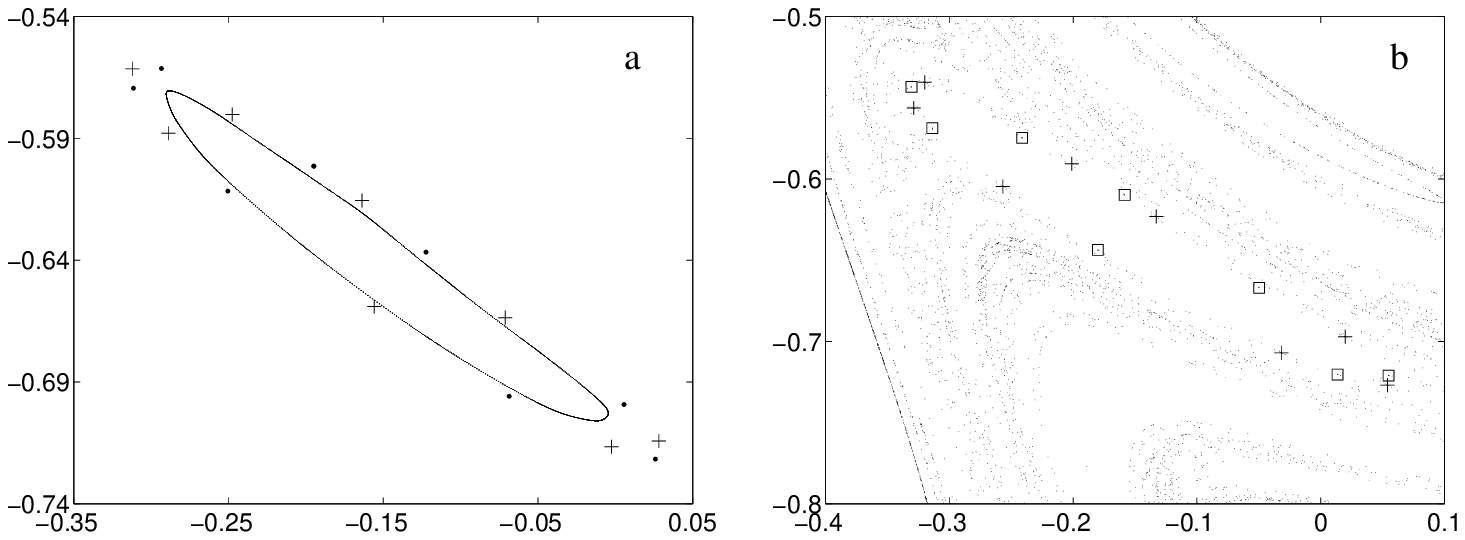,width=400pt}}
\end{picture}
\scaption{Poincar\'e sections of the Lorenz-84 model with $a=1/4$ and $b=4$. a: $(F,G)=(8,0.9725)$.
Two coexisting
invariant tori, one with the period 8 orbits, one with quasiperiodic dynamics. The thick dots denote
the stable orbit, the crosses the saddle type orbit.
b: $(F,G)=(8,0.99)$.
Beyond the
torus and flip bifurcations $\mbox{TR}$ and $\mbox{F}$. Both orbits, marked with crosses and boxes
are now of saddle type.}
\label{secs}
\end{figure}

The phase angle crosses the point $\phi=3\pi/4$ twice, at resonance points $\mbox{A}_{1,2}$.
One edge of the resonance tongue connects these points.
It appears that the invariant torus itself goes through two subsequent fold bifurcations,
so that the edge of the resonance tongue can cross $\mbox{tr}_{2}$, and in a narrow band
two stable tori coexist. This is shown in figure (\ref{secs}a). On the inner torus the dynamics
is quasiperiodic, on the outer torus the period 8 orbits exist. If we further increase $G$, the
period 8 orbit loses its stability in a torus bifurcation, directly followed by a flip bifurcation,
as one of the multipliers crosses back into the unit circle through $-1$. Beyond these bifurcations,
a chaotic attractor appears. The corresponding Poincar\'e section is shown in figure (\ref{secs}b).
The whole picture is more involved, as the tori shown in figure \rf{secs}a coexist with several
tori with phase locked orbits of higher period.

Scenarios for the creation of a chaotic attractor through the bifurcation of a periodic orbit
on on invariant torus are described in \cite{bro2}. The Ruelle-Takens scenario in the vicinity of an 
angular degeneracy was recently found in an electronic model by \cite{alga}.

\subsubsection{Shil'nikov bifurcations}

The neutral saddle focus transitions, found in the six dimensional model, \rf{odesys1}, and the 
approximate reduced model, \rf{odesys2}, are not present in diagrams \rf{diagram3}-\rf{diagram5}.
Therefore, no Shil'nikov type chaos occurs in the Lorenz-84 model for these parameter values.
We can, however, retrace the transition points for different values of $b$. As the continuation 
package HomCont (\cite{doed}) allows for three parameter continuation of codimension two points on 
homoclinic curves,
it is possible to calculate a curve of neutral saddle focus transitions in the space of parameters
$b$, $F$ and $G$. It turns out, that two transition points appear on curve $\mbox{H}_{1}$ if
$b<0.419$. Just like in diagram \rf{diagram2}, on a small segment the saddle value becomes
negative. Therefore, for $b<0.419$ Shil'nikov type chaos can be encountered in the Lorenz-84 model.

\subsubsection{Intermittency}

The last route to chaos described here is through intermittency. If we keep $F=8$ fixed and
increase $G$, we find a periodic window around $G\approx 1.167$. At $G=1.16742$, the stable
periodic orbit undergoes a saddle node bifurcation. Beyond this point, the behaviour is
intermittent. In terms of \cite{pome}, this is type I intermittency. Two time series in the
intermittent regime are shown in figure \rf{interm}.
\begin{figure}[t]
\begin{picture}(400,150)
\centerline{\epsfig{file=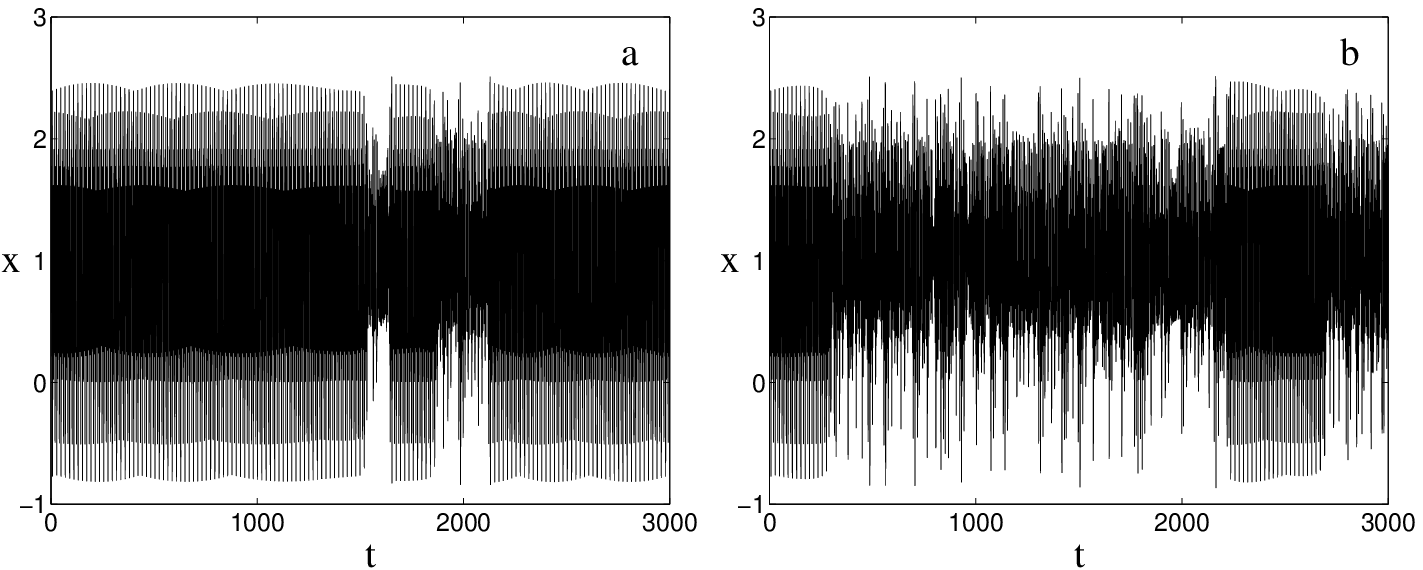,width=400pt}}
\end{picture}
\scaption{Intermittency in the Lorenz-84 model. a: $(F,G)=(8,1.16743)$. b: $(F,G)=(8,1.1675)$.}
\label{interm}
\end{figure}
In \cite{wiec} the same
type of intermittency is found and pictures of the stable and unstable manifolds of the saddle
type orbit are shown.

\section{Conclusion}

The starting point of the analysis in this paper is the truncation to six degrees of freedom
of a QG two level model of atmospheric flow. Period doubling cascades and Shil'nikov
bifurcations are identified as routes to chaos in the bifurcation diagram of this model.
A measurement of the dimension of the chaotic attractor along a section of the bifurcation
diagram reveals that it is less than three dimensional, hinting at the existence of a three
dimensional invariant manifold which captures the dynamics.

In order to approximate this invariant manifold we have introduced a small parameter, $\epsilon$,
into the equations. In the limit of $\epsilon\downarrow 0$ an exact solution was presented.
This solution has a clear physical interpretation in terms of energy exchange between the
zonally symmetric jet stream mode and the traveling waves. Numerical evidence for the existence
of this manifold for finite $\epsilon$ was found implementing a variant of the graph transform algorithm
of \cite{bro1}.

We have shown that we can set $\epsilon=0$ without destroying the qualitative dynamics.
The bifurcation diagram of the reduced model in this limit agrees well with the diagram of
the six dimensional model. The invariant manifold at $\epsilon=0$ can be linearised around the
Hadley state. If the six dimensional model is
reduced to this linear approximation of the invariant manifold, the Lorenz-84 model is found.
The parameters of the Lorenz-84 model can then be calculated from the physical parameters of
the QG model. One of them comes out significantly different. We have compared the bifurcation
diagram of the Lorenz-84 model at physical parameters to that of the six dimensional model and
to the diagram at traditional parameter values, first presented in \cite{shil}.

Finally, we have discussed four possible routes to chaos in the Lorenz-84 model. Period
doubling cascades appear both in the six dimensional model and in the Lorenz-84 model.
Shil'nikov bifurcations are only found in the Lorenz-84 model for parameter values far
away from those considered here. A route to chaos not considered before in this model
is the Ruelle-Takens scenario. We have presented evidence that the creation and destruction
of invariant tori and the presence of resonance tongues lead to chaos in the Lorenz-84 model.
Finally, an intermittent transition has been presented. The overall picture of chaotic
attractors being created and destroyed via different routes is reminiscent of the dynamics
described in \cite{wiec}.

The link between a Galerkin truncation of a QG baroclinic model and the Lorenz-84 model
justifies the use of the latter in conceptual studies of atmosphere and climate dynamics.
It is remarkable how much of the bifurcation structure of the six dimensional truncation
is preserved, notwithstanding rough approximations (namely setting $\epsilon=0$
and linearising the three dimensional invariant manifold). Probably, this is because the
six dimensional model isolates one aspect of midlatitude, synoptic flow: the energy
exchange between the jet stream and the baroclinic waves. When reducing to three degrees
of freedom, and subsequently to the Lorenz-84 model, this process is modeled qualitatively 
correctly.

\section{Acknowledgments}

The author wishes to thank Ferdinand Verhulst, Theo Opsteegh, Ulrich Achatz, Yuri Kuznetsov
and Renato Vitolo for useful discussions and proofreading.

\clearpage
\bibliography{IJBC}
\bibliographystyle{IJBC}
\end{document}